\documentclass[onecolumn]{aastex631}
\usepackage{url}
\usepackage{multirow}
\usepackage{amsmath}

\begin{document}

\title{The pre-perihelion evolution of the activity of comet C/2017 K2 (PANSTARRS) during the water ice-line crossover\footnote{Based on observations made with ESO Telescopes at the Paranal Observatory under program 109.23D6.001 (PI: Y. G. Kwon)}}
\shorttitle{The activity evolution of C/2017 K2}

\received{\today}
\revised{---}
\accepted{---}

\author[0000-0002-8122-3606]{Yuna G. Kwon}
\affiliation{Caltech/IPAC, 1200 E California Blvd, MC 100-22, Pasadena, CA 91125, USA}
\affiliation{Institut f{\" u}r Geophysik und Extraterrestrische Physik, Technische Universit{\"a}t Braunschweig,  \\
Mendelssohnstr. 3, 38106 Braunschweig, Germany}

\author{Stefano Bagnulo}
\affiliation{Armagh Observatory, College Hill, Armagh BT61 9DG, Northern Ireland, UK}

\author[0000-0001-5989-3630]{Johannes Markkanen}
\affiliation{Institut f{\" u}r Geophysik und Extraterrestrische Physik, Technische Universit{\"a}t Braunschweig,  \\
Mendelssohnstr. 3, 38106 Braunschweig, Germany}

\author{Ludmilla Kolokolova}
\affiliation{Department of Astronomy, University of Maryland, College Park, MD 20742, USA}

\author{Jessica Agarwal}
\affiliation{Institut f{\" u}r Geophysik und Extraterrestrische Physik, Technische Universit{\"a}t Braunschweig,  \\
Mendelssohnstr. 3, 38106 Braunschweig, Germany}
\affiliation{Max Planck Institute for Solar System Research, Justus-von-Liebig-Weg 3, 37077 Göttingen, Germany}

\author{Manuela Lippi}
\affiliation{INAF - Osservatorio Astrofisico di Arcetri - Largo Enrico Fermi, 5, 50125 Firenze, Italy}

\author[0000-0002-6610-1897]{Zuri Gray}
\affiliation{Armagh Observatory, College Hill, Armagh BT61 9DG, Northern Ireland, UK}
\affiliation{Mullard Space Science Laboratory, University College London, Holmbury St. Mary, Dorking RH5 6NT, UK}

\correspondingauthor{Yuna G. Kwon}
\email{yuna@ipac.caltech.edu}

\begin{abstract}

Comets, relics from the early solar system, consist of dust and ice. The ice sublimates as comets approach the Sun, ejecting dust from their nuclei seen as activity. Different volatiles sublimate at different Sun-comet distances and eject dust of unique sizes, structures, and compositions. In this study, we present new polarimetric observations of Oort-cloud comet C/2017 K2 (PANSTARRS) in R and I-filter domains before, during, and after its crossover of the water-ice sublimation regime at phase angles of 15.9\arcdeg, 10.5\arcdeg, and 20.0\arcdeg, respectively. Combining multiband optical imaging data covering a wide range of heliocentric distances ($\sim$14$-$2.3 au), we aim to characterize the preperihelion evolution of cometary activity as well as the properties of its coma dust. Two discontinuous brightening events were observed: at $\sim$6 au presumably associated with changes in CO-like supervolatile ice activity, and at $\sim$2.9 au when water ice took over. Particularly, the latter activation is accompanied by changes in coma morphology and color whose trends differ between the inner ($\sim$10$^3$-km) and outer ($\sim$10$^4$-km) parts of the coma. No polarimetric discontinuities on the comet were observed over the inner coma region, all epochs showing phase-angle and wavelength dependencies compatible with those of active comets observed in similar observing geometry. During this period, the underlying dust continuum overwhelmed H$\alpha$ emission at around 656.3 nm, suggesting less water ice on the comet's surface than expected. We discuss K2's coma environment by combining numerical simulations of light scattered by dust and place the observations within the context of the comet's evolution.

\end{abstract}


\section{Introduction} \label{sec:intro}

Comets shed large amounts of dust during the approach to the Sun that is visible as cometary comae and dust tails. 
Activity mechanisms are strongly influenced by temperature or heliocentric distance. Sublimation of supervolatile ices (e.g., CO or CO$_2$) and exothermic crystallization of amorphous ice dominate activity in giant planet regions, while sublimation of water ice starts around the orbit of Jupiter ($\sim$5 au) and dominates activity within $\sim$3 au \citep{Prialnik2004,Meech2009,Blum2014,Womack2017}. Each mechanism ejects dust in distinctive sizes, structures, and possibly compositions (e.g., \citealt{Gundlach2015,Gundlach2020}), hence characterizing a change in the style of mass loss as the comet passes by the critical heliocentric distance is vital to diagnosing the dust nature that makes up the comet nucleus' near-surface layer \citep{Fulle2022}.

Comet C/2017 K2 (PANSTARRS) (hereafter K2) is an Oort-cloud comet on its way to perihelion, crossed the so-called water ice line ($\sim$2.7 au; \citealt{Blum2014,Womack2017}) in July 2022. With a dust production rate nearly 100$–$1,000 times greater than most comets at similar distances (e.g., \citealt{Licandro2019,Garcia2020}), K2 presents a unique opportunity to probe an activity regime previously not observed in detail. The unprecedented level of activity that it was showing even outside the orbit of Saturn suggests that dust production may have begun in the Kuiper Belt \citep{Jewitt2021}. Its supervolatile ice-driven activity was also supported by the detection of its CO emission line at $\sim$6.7 au \citep{Yang2021} and its perpetual activity observed beyond 4 au \citep{Milewski2024}. Polarimetric studies made at similar heliocentric distances ($\sim$6.8$-$7.1 au) also display measurable level of cometary activity \citep{Zhang2022,Kochergin2023}.
The inner solar system activity of K2 was recently demonstrated by \citet{Kwon2023a} by showing distinct dust populations with different ejection times and inhabiting different regions in its coma. This single-epoch observation at $\sim$2.53 au from the Sun with the Multi Unit Spectroscopic Explorer (MUSE) suggests that the comet's activation status has changed during its inbound leg \citep{Kwon2023a}.

This paper presents new imaging polarimetric observations of K2 dedicated to studying the evolution of its preperihelion activity by covering before, during, and after the crossover of the region at around 2.7 au. 
The use of polarimetry can offer directional information about the light scattered by cometary dust particles resulting from electromagnetic interactions between incident sunlight and dust constituents to understand the microscale physics of cometary dust particles \citep{Bohren1983,Mishchenko2010}. Multiband optical images from Zwicky Transient Facility (ZTF; \citealt{Bellm2019,Graham2019}) obtained over a wide range of heliocentric distances and a single-epoch infrared image from Near-Earth Object Wide-field Infrared Survey Explorer (NEOWISE; \citealt{Mainzer2011}) taken outside the orbit of Jupiter, along with numerical simulations of the interaction of light with dust particles, will help us interpret our polarimetric observations in the broader context of comet activity evolution.
\\

\section{Observations and data reduction} \label{sec:obsdata}

The main observational data of K2 in this study were obtained using the 8.0-m diameter Very Large Telescope (VLT) in Chile. Following is a summary of the optics used, observing conditions, and data analysis methodology. 

\subsection{Observations} \label{sec:obs}

A three-epoch observation of K2 in imaging and imaging polarimetry was conducted from May to July 2022 to cover the water ice-line passage of the comet. A multi-purpose optical instrument (FORS2; \citealt{Appenzeller1998}) mounted on the Cassegrain focus of the UT1 telescope at VLT at the Paranal Observatory (70$\arcdeg$24$\arcmin$10$\arcsec$ W 24$\arcdeg$37$\arcmin$31$\arcsec$S, 2 635 m) has been used for observation. FORS2 is equipped with imaging and dual-beam polarimetric optics and their detailed specifications and principles of operation are described in the FORS2 User Manual\footnote{\url{https://www.eso.org/sci/facilities/paranal/instruments/fors/doc/VLT-MAN-ESO-13100-1543\_P112.pdf}} and references therein.

Our observation runs were carried out under the program (ID: 109.24F3.001), each of which consists of one observing block of H$\alpha$-band imaging (with the H\_Alpha+83 filter at 656.3 $\pm$ 6.1 nm) and two blocks of intermediate-band imaging polarimetry (with the FILT\_691\_55+69 and FILT\_834\_48+71 continuum filters at (691 $\pm$ 55) nm and (834 $\pm$ 48) nm, respectively). The first block aims to cover the H$\alpha$ emission band at $\sim$656.3 nm, while the filters in the latter block were selected to monitor the secular evolution of dust activity of K2 minimizing possible contamination by gas molecules. All setups implemented a standard resolution (SR) collimator resulting in an image scale of 0.25$\arcsec$/pixel in the default 2$\times$2 binned, low-gain read-out mode. For H$\alpha$ imaging observations, we made 7 exposures with a single exposure time of 30 secs. For each continuum filter in linear polarimetry, we obtained two sets of rotatable half-wave plate (HWP) orientations ranging from 0\arcdeg\ to 157.5$\arcdeg$ separated by 22.5$\arcdeg$ with a Wollaston prism (Woll\_34+13). Identical instrument configurations were employed for all three observing epochs, where median seeing was $\sim$0.82, $\sim$0.62, and $\sim$0.68 for the May, June, and July observations, respectively. A pixel's projected physical distance from the comet nucleus at each epoch is $\sim$475 km, $\sim$336 km, and $\sim$335 km, respectively, in default 2$\times$2-binned mode.
Table \ref{t01} summarizes the observing circumstances and K2 geometry.

\begin{deluxetable*}{c|c|c|c|c|c|c|c|cccc}
\tabletypesize{\footnotesize} 
\tablewidth{0pt} 
\tablecaption{Journal of observations of C/2017 K2 (PANSTARRS) \label{t01}}
\tablehead{
\colhead{Telescope}  \vline& 
\colhead{\multirow{2}{*}{Name}}  \vline&
\colhead{UT Date}  \vline& 
\colhead{\multirow{2}{*}{Mode}}  \vline&
\colhead{\multirow{2}{*}{Filter}}  \vline& 
\colhead{\multirow{2}{*}{$N$}}  \vline&
\colhead{Exptime}  \vline& 
\colhead{\multirow{2}{*}{Airmass}}  \vline&
\colhead{$r_{\rm H}$} & 
\colhead{$\Delta$}  &
\colhead{$\alpha$}  &
\colhead{$\nu$}\\
Instrument & & (Year 2022) & & & & (sec) & & (au) & (au) & ($\arcdeg$) & ($\arcdeg$)
}
\startdata
\multirow{16}{*}{VLT-UT1/} & \multirow{6}{*}{Epoch 1} & \multirow{6}{*}{May 10 06:40} & \multirow{2}{*}{Img} & \multirow{2}{*}{H$\alpha$} & \multirow{2}{*}{7} & \multirow{2}{*}{30} & 1.47 & \multirow{6}{*}{3.227} & \multirow{6}{*}{2.617} & \multirow{6}{*}{15.9} & \multirow{6}{*}{276.5} \\
\multirow{17}{*}{FORS2} & & & & & & & (1.48--1.46) & & & \\
\cline{4-8}
 & & & \multirow{4}{*}{Impol} & \multirow{2}{*}{FILT\_691\_55} & \multirow{4}{*}{16} & \multirow{4}{*}{60} & 1.38 & & & &  \\
 & & & & & & & (1.42--1.34) & & & & \\
 & & & & \multirow{2}{*}{FILT\_834\_48} & & & 1.31 & & & &  \\
 & & & & & & & (1.34--1.28) & & & & \\
\cline{2-12}
 & \multirow{6}{*}{Epoch 2} & \multirow{4}{*}{June 29 05:30} &  \multirow{4}{*}{Impol} & \multirow{2}{*}{FILT\_691\_55} & \multirow{4}{*}{16} & \multirow{4}{*}{60} & 1.23 & \multirow{4}{*}{2.783} & \multirow{4}{*}{1.855} & \multirow{4}{*}{10.5} & \multirow{4}{*}{287.0} \\
 & & & & & & & (1.20--1.26) & & & & \\
 & & & & \multirow{2}{*}{FILT\_834\_48} & & & 1.32 & & & &  \\
 & & & & & & & (1.27--1.37) & & & & \\
\cline{3-12}
 & & \multirow{2}{*}{July 01 03:11$^{\star}$} & \multirow{2}{*}{Img} & \multirow{2}{*}{H$\alpha$} & \multirow{2}{*}{7} & \multirow{2}{*}{30} & 1.12 & \multirow{2}{*}{2.767} & \multirow{2}{*}{1.843} & \multirow{2}{*}{10.8} & \multirow{2}{*}{287.4} \\
 & & & & & &  & (const.) & & & & \\
\cline{2-12}
 & \multirow{6}{*}{Epoch 3} & \multirow{6}{*}{July 29 00:27} & \multirow{2}{*}{Img} & \multirow{2}{*}{H$\alpha$} & \multirow{2}{*}{7} & \multirow{2}{*}{30} & 1.05 & \multirow{6}{*}{2.530} & \multirow{6}{*}{1.849} & \multirow{6}{*}{20.0} & \multirow{6}{*}{294.9} \\
 & & & & & & & (const.) & & & \\
\cline{4-8}
 & & & \multirow{4}{*}{Impol} & \multirow{2}{*}{FILT\_691\_55} & \multirow{4}{*}{16} & \multirow{4}{*}{60} & 1.03 & & & &  \\
 & & & & & & & (1.04--1.03) & & & & \\
 & & & & \multirow{2}{*}{FILT\_834\_48} & & & 1.03 & & & &  \\
 & & & & & & & (const.) & & & &
\enddata
\tablecomments{Top headers: Mode, `Img' and `Impol' for imaging and imaging polarimetric observation modes, respectively; $N$, number of exposures; Exptime, total integration time in seconds; $r_{\rm H}$ and $\Delta$, mid-point heliocentric and geocentric distances over a nightly observation in au, respectively; $\alpha$, mid-point phase angle (angle of Sun$-$comet$-$observer) in degrees; $\nu$, mid-point true anomaly in degrees. \\
$^\star$The H$\alpha$ imaging observation was initially made on the same date as imaging polarimetric observations (on UT 2022 June 29). However, since a photometric standard star in H$\alpha$ was not observed during the observing night, we conducted additional observation on UT 2022 July 01.}
\end{deluxetable*}

\subsection{Data analysis} \label{sec:data}

All raw fits files were preprocessed according to a standard procedure (i.e., bias subtraction, flat-fielding, and sky subtraction). H$\alpha$ images were median-combined on the World Coordinate System information of K2 at the time of observations using \texttt{astropy.wcs}\footnote{\url{https://docs.astropy.org/en/stable/wcs/}}. The remaining ill-behaved pixels hit by cosmic rays were corrected using the \texttt{lacosmic} python package\footnote{\url{https://lacosmic.readthedocs.io/en/stable/}} whose original algorithm is described in \citet{vanDokkum2001}.

FORS2, as a dual-beam polarimeter, consists of a rotatable HWP and Wollaston prism, providing an intensity image of polarized light elements measured at different HWP angles $\phi$ (0\arcdeg, 22.5\arcdeg, 45\arcdeg, and 67.5\arcdeg). We implemented the so-called beam-swapping technique to correct the first-order instrumental effects, such as chromatism of the retarder waveplate (e.g., \citealt{Bagnulo2009}), following the calibration and extraction of polarimetric parameters in the same way as \citet{Bagnulo2015,Bagnulo2023}. Reduced Stokes parameters $P^\prime_{Q}$ (=$Q$/$I$) and $P^\prime_{U}$ (=$U$/$I$) on two sets of Stokes parameters $Q$ and $U$ in units of percent can be measured in the instrument reference system by 
\begin{equation}
P^\prime_{X} = \frac{1}{2{\rm n}}\Bigg\{\Bigg(\frac{f_{o} - f_{e}}{f_{o} + f_{e}}\Bigg)_{\phi_{X}} - \Bigg(\frac{f_{o} - f_{e}}{f_{o} + f_{e}}\Bigg)_{\phi_{X} + 45\arcdeg}\Bigg\} - P_{X, {\rm inst}}
\label{eq:eq1}
\end{equation}
\noindent where $X$ denotes either $Q$ or $U$ and $f_{o}$ and $f_{e}$ are photon counts of ordinary (parallel) and extraordinary (perpendicular) beams observed at $\phi$. n is the number of exposure pairs ($n$ = 2 in our case), each set consisting of exposures at four values of $\phi$. Images of $\phi_{Q}$ $\in$ \{0\arcdeg, 90\arcdeg\} and $\phi_{U}$ $\in$ \{22.5\arcdeg, 112.5\arcdeg\} were used for $P^\prime_{Q}$ and $P^\prime_{U}$ estimations, respectively. $P_{X, {\rm inst}}$ denotes the instrumental linear polarization degree adopted from \citet{Cikota2017}.
Outputs of Eq. \ref{eq:eq1} were then transformed into the reference direction (normal to the scattering plane which is the plane passing through the Sun$-$comet$-$observer) using the equation of
\begin{equation}
\begin{split}
P_{Q} & = P^\prime_{Q} \cos(2\theta) + P^\prime_{U} \sin(2\theta)\\
P_{U} & = -P^\prime_{Q} \sin(2\theta) + P^\prime_{U} \cos(2\theta)~,
\end{split}
\label{eq:eq2}
\end{equation}
\noindent where $\theta$ = \texttt{ADA.POSANG} + $\epsilon$ + $\Phi$ + $\pi$/2 is the angle of rotation applied. A header keyword \texttt{ADA.POSANG} specifies the position angle on the sky and thus the amount of rotation of the instrument position angle so that the x and y axes of the field of view are aligned with the East and North directions of the source. This parameter in all of our observations was set to zero. $\epsilon$ is the instrumental polarization angle induced by the wavelength-dependent chromatism of the HWP. This value is provided by the FORS2 official webpage\footnote{\url{https://www.eso.org/sci/facilities/paranal/instruments/fors/inst/pola.html}}, from which we applied an $\epsilon$ value of $-$2.75\arcdeg\ and $-$2.76\arcdeg\ to the FILT\_691\_55 and FILT\_834\_48 data, respectively. $\Phi$ is the position angle of the scattering plane whose sign was chosen to satisfy 0 $\leq$ ($\Phi$ $\pm$ 90\arcdeg) $\leq$ 180\arcdeg. As a result, the degree of linear polarization of cometary dust $P_{\rm r}$ and its position angle $\theta_{\rm r}$ are
\begin{equation}
\begin{split}
P_{\rm r} & \simeq P_{Q}\\
\theta_{\rm r} & = \frac{1}{2}{\rm atan2}\Bigg(\frac{P_{U}}{P_{Q}}\Bigg)~.
\end{split}
\label{eq:eq3}
\end{equation}
\noindent
The \texttt{atan2} function was adopted instead of \texttt{arctan} to resolve $\pi$-ambiguity of the polarization vectors \citep{Bagnulo2006}. When $\theta_{\rm r}$ = 0\arcdeg, that is, the angle perfectly aligns perpendicular to the scattering plane (subscript $r$ denotes reference), $P_{Q}$ = $P_{\rm r}$ and $P_{U}$ = 0 \%. When $P_{U}$ values are significantly deviating from zero, the accuracy of any detected features is less reliable; therefore, for data points only with moderately small $P_{U}$ ($<$1 \% in absolute numbers, otherwise rejected; Sect. \ref{sec:res1}), we considered $P_{Q}$ as an equivalent to $P_{\rm r}$ and used both quantities interchangeably. In addition, Null parameters $N_{Q}$ and $N_{U}$ in the $P_{Q}$ and $P_{U}$ spaces, respectively, were adopted for quality checks on polarimetric results (e.g., \citealt{Bagnulo2023}), whose equations and spatial distributions on the polarization maps are provided in Appendix \ref{sec:app1}. For any measured polarimetric features (i.e., their values and distributions across the coma) having either $|$$N_{Q}$$|$ and $|$$N_{U}$$|$ of $\gtrsim$1.0 \%, we flagged the results and excluded them from consideration. All $P_{\rm r}$ (thus $P_{Q}$) values introduced in the following sections have null parameters that cluster around 0 \% within $\sim$8--10\arcsec-radius coma areas around the photocenter. 

The intermediate-band filters we implemented in polarimetry are located outside the major emission bands of gas molecules at visible wavelengths \citep{Meech2004}; thus, we anticipate that any polarimetric results we present below will have negligible gas impacts. This treatment would possibly be supported by the visible spectra of K2 acquired around the inner boundary of the water ice line lacking any significant emission features within the wavelength range of interest  (Fig. 2 in \citealt{Kwon2023a}).
\\

\section{Results} \label{sec:res}

\begin{figure}[bh]
\centering
\includegraphics[width=0.5\textwidth]{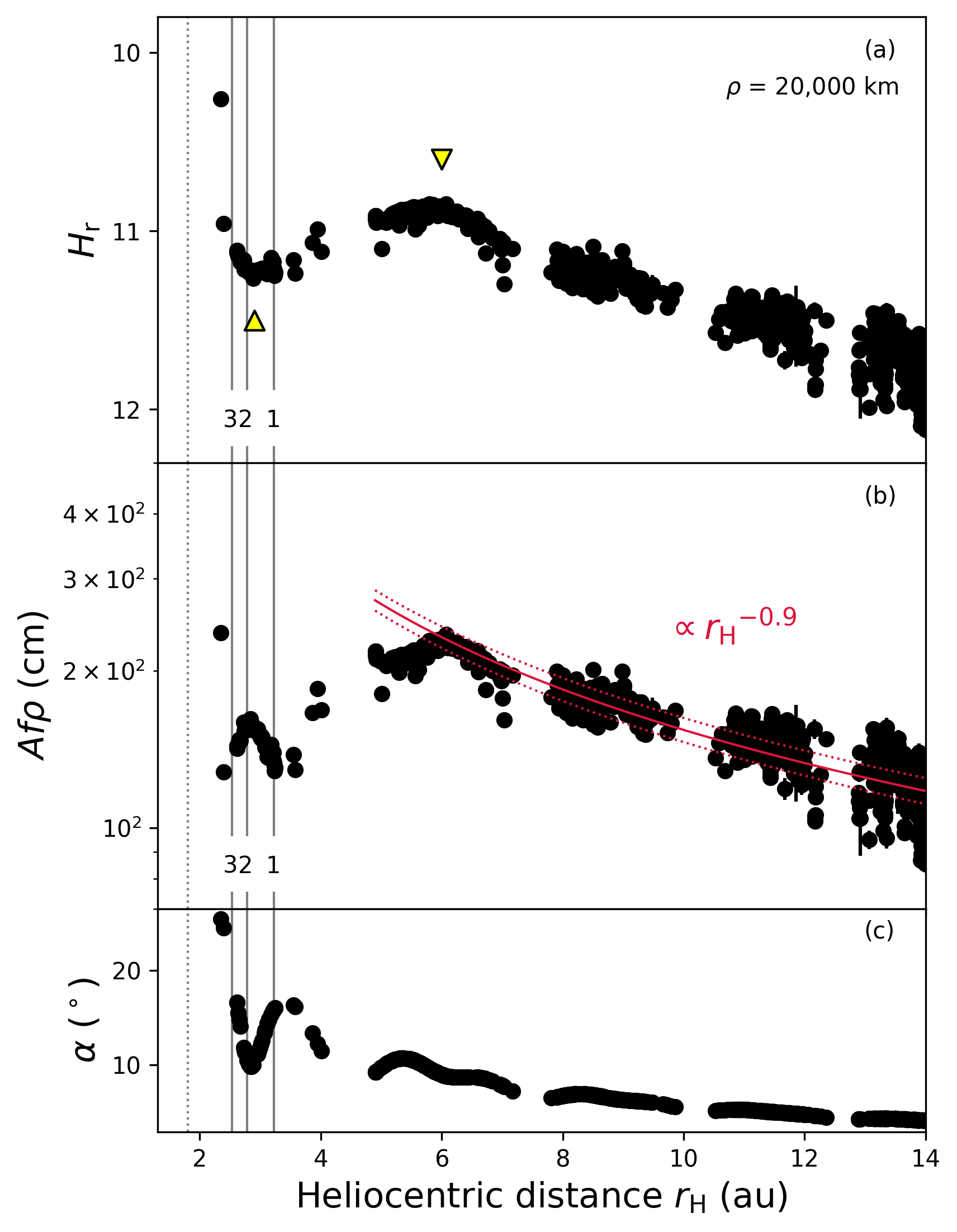}
\caption{Secular evolution of $r$-band brightness and {\it Af$\rho$} parameter of K2 as a function of heliocentric distance $r_{\rm H}$. The $r$-band absolute magnitude $H_{\rm r}$ corrected for the heliocentric and geocentric distances and phase effects (Eq. \ref{eq:eq0}) is presented in panel a, while {\it Af$\rho$} parameter was calculated from Eq. \ref{eq:eq0-2} and is shown in panel b. Panel c illustrates the impact of an abrupt change in K2's $\alpha$ during our heliocentric coverage.
All points correspond to the photometric parameters measured from the aperture size $\rho$ of 20,000 km from the photocenter. Vertical lines indicate the $r_{\rm H}$ of our VLT observations at three epochs (Table \ref{t02}). Two inflection points where K2 experienced a change in its $r$-band brightness are marked with yellow triangles in panel a. 
\label{Fig01}}
\end{figure}

Figure \ref{Fig01} summarizes the secular evolution of K2's $r$-band brightness and {\it Af$\rho$} parameter, which is a proxy for dust mass-loss rates in cometary comae \citep{A'Hearn1984}, across a range of heliocentric distances $r_{\rm H}$ from $\sim$14 to $\sim$2.3 au in its inbound leg using the Zwicky Transient Facility (ZTF) IRSA archive\footnote{\url{https://irsa.ipac.caltech.edu/applications/ztf/?__action=layout.showDropDown&}} \citep{Masci2018}. 

For correction of varying viewing geometry and investigation of the evolution of K2's intrinsic $r$-band brightness, the apparent $r$-band magnitude (in the zr-band filter with effective wavelength 636.7 nm and FWHM 155.3 nm) was converted into absolute magnitude $H_{\rm r}$, which corresponds to a hypothetical solar-center magnitude ($r_{\rm H}$ = $\Delta$ = 1au and $\alpha$ = 0\arcdeg), using
\begin{equation}
H_{\rm r} = m_{\rm r,app} - 5\log(\Delta r_{\rm H}) - 2.5\log(\Phi(\alpha))~,
\label{eq:eq0}
\end{equation}
\noindent whose results are shown in Figure \ref{Fig01}a. Here, $m_{\rm r,app}$ is the apparent $r$-band magnitude measured with an aperture radius covering the cometocentric distance $\rho$ of 20,000 km and $\Phi(\alpha)$ is the phase function for correction of the phase darkening of the dust cloud observed at the phase angle $\alpha$ to $\alpha$ = 0\arcdeg. The aperture radius of $\rho$ = 20,000 km was selected for our photometric analysis because 1) it is the size that well encapsulates the large-scale coma information (e.g., \citealt{Marschall2020b}), and 2) it is larger than FWHMs of all observing epochs in a range of geocentric distances, yet small enough to avoid severe background star overlap. 
Due to the lack of studies on phase curves on K2, we adopted a commonly used empirical phase function, 2.5$\log$($\Phi(\alpha)$) = $\beta\alpha$ with $\beta$ = 0.035 mag deg$^{-1}$ assumed \citep{Lamy2004}.
The $m_{\rm r,app}$ was also used to estimate the {\it Af$\rho$} parameter (panel b in Fig. \ref{Fig01}) by
\begin{equation}
Af\rho = {\rm Y}\Bigg[\frac{\Delta}{\rm au}\Bigg]^{2}\Bigg[\frac{r_{\rm H}}{\rm au}\Bigg]^{2}\Bigg[\frac{\rho}{\rm cm}\Bigg]^{\rm -1} \times 2.512^{(m_{\rm \sun,r} - m_{\rm r,app})}
~,
\label{eq:eq0-2}
\end{equation}
\noindent where Y is a distance unit conversion factor of 8.95 $\times$ 10$^{26}$ (Eq. 8 in \citealt{Kwon2019}) with $\rho$ in centimeters (2 $\times$ 10$^9$ cm). $m_{\rm \sun,r}$ = $-$27.04 is the apparent $r$-band magnitude of the Sun \citep{Willmer2018}. Figure \ref{Fig01}c shows the phase angle ($\alpha$) change over the term. As {\it Af$\rho$} uses $m_{\rm r,app}$, which is not corrected for phase changes, the rapid shift in $\alpha$ between 2.5 and 3.5 au caused the {\it Af$\rho$} to fluctuate in the opposite direction.

Even with a single exposure time of 30 sec implemented in the ZTF, K2 displayed bright coma signals well outside Saturn's orbit due to abnormally high coma activity. It is thus possible to observe K2 while having passed through several regimes with cometary activity driven by different types of ice (primarily CO, CO$_2$, and H$_2$O from far to close; see e.g., Table 2 of \citealt{Fulle2022} for a summary). The intrinsic brightness of K2 (as represented by $H_{\rm r}$) gradually increases from 14 au to $\sim$6 au, then stagnates or slightly decreases between 3 and 6 au, before rapidly brightening again near 2.9 au as the comet approaches the Sun. In Figure \ref{Fig01}a, two inflection points (at $\sim$2.9 au and 6 au) are marked where the heliocentric evolution of the absolute magnitude $H_{\rm r}$ changes its slope. The heliocentric evolution of {\it Af$\rho$} outside the outer inflection point at $r_{\rm H}$ $>$ 6 au follows a power-law function with an index of $\sim$0.9. Three vertical lines in Figure \ref{Fig01} indicate our VLT observations (Table \ref{t01}). Considering that the triangle near Epoch $\sharp$2 is located around the expected launch distance of full-scale water ice-driven activity \citep{Blum2014,Womack2017}, our Epochs $\sharp$1, $\sharp$2, and $\sharp$3 VLT measurements may approximate the moments of K2 before, around, and after the serious water ice-driven activation that likely occurred at $\sim$2.9 au from the Sun. In Sect. \ref{sec:dis}, we will discuss further implications of the observed trends.
\\

\subsection{Polarization maps} \label{sec:res1}

\begin{figure*}[t]
\centering
\includegraphics[width=\textwidth]{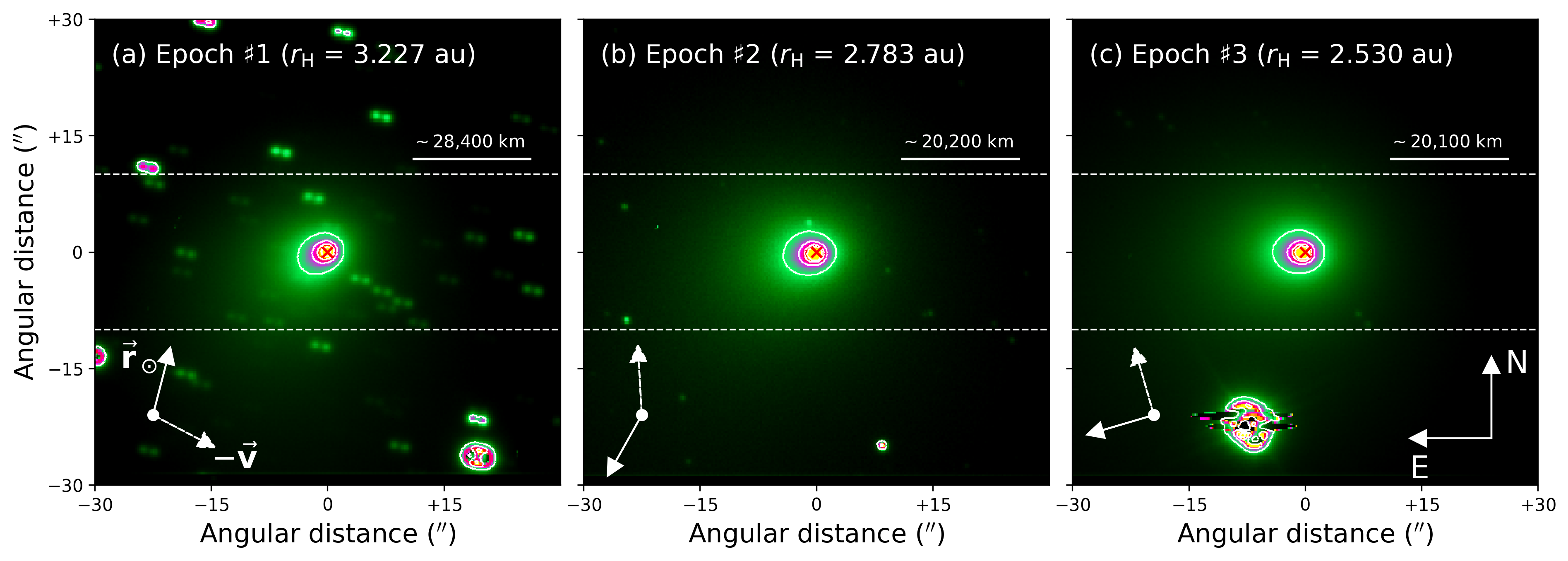}
\caption{60-sec-exposure dust acquisition images of K2 in the Red domain (FILT\_691\_55) with VLT/FORS2 at three epochs of observations. Two dashed horizontal lines bracket the sky area covered by the polarization strip. $\vec{\bf{r}}_\sun$ and $-\vec{\bf{v}}$ denote the anti-sunward vector and the negative velocity direction of the comet, respectively. Photocenters are marked with a red `X' symbol in the center. Contours represent 90, 70, 50, 30, and 10 \%\ of the brightness peak.
\label{Fig02}}
\end{figure*}

Figure \ref{Fig02} shows the acquisition images of the K2 dust coma in the FILT\_691\_55 filter (Red domain) taken at three observing epochs right before the main polarization observation block.
These images are for checking the coma brightness distribution along the polarization strip and thus examining macroscopic coma features, not for photometric analysis due to their shallowness: two images were stacked for the first two epochs and three images for the last one. With $\sim$10 \% of the peak dust coma signal reached $\sim$4$-$5\arcsec\ (i.e., about one-third of the width of the polarization strip), no significant anisotropy was detected. 

\begin{figure*}[ht]
\centering
\includegraphics[width=0.77\textwidth]{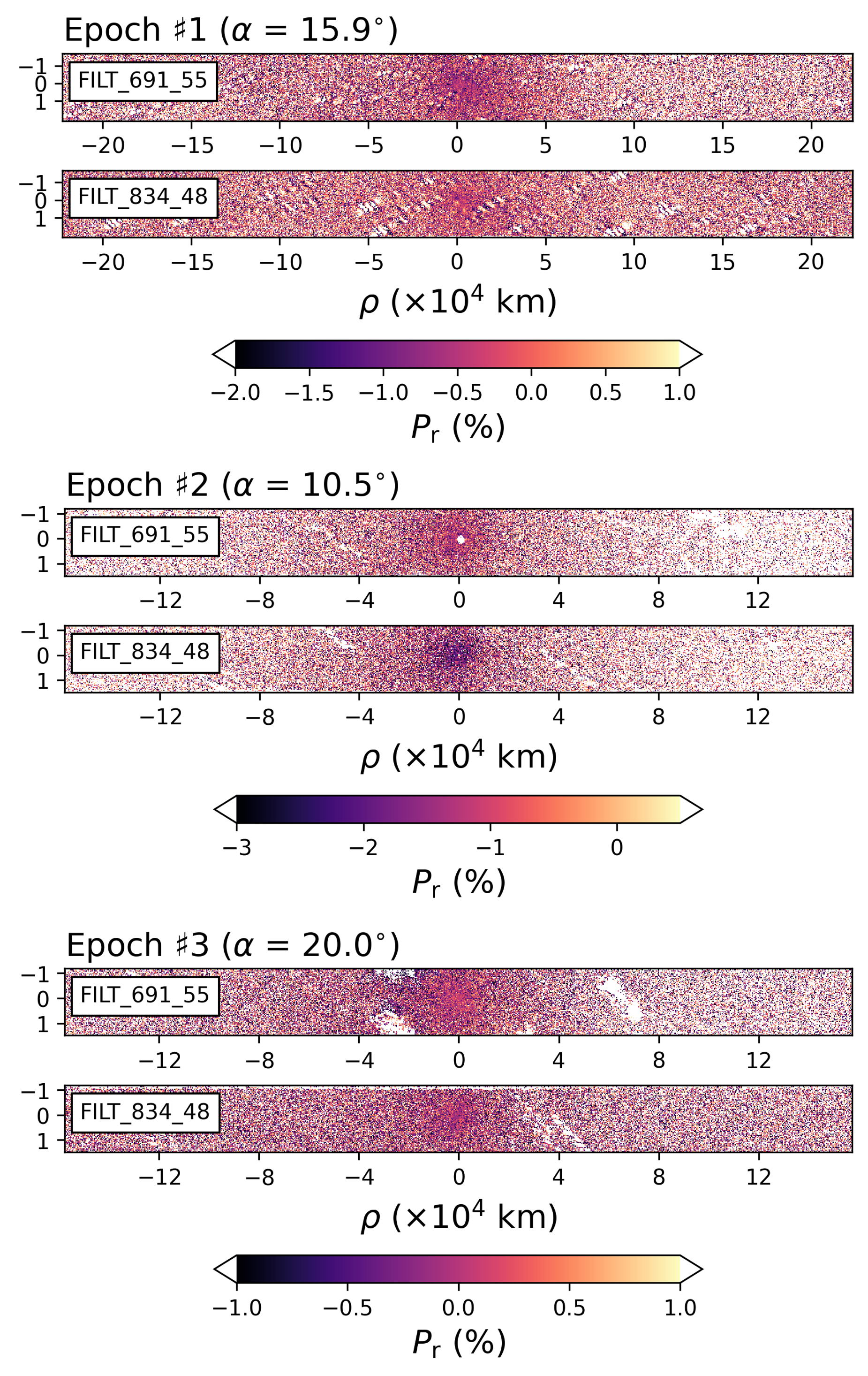}
\caption{Polarization maps of K2. Rows represent maps in each filter at a particular epoch. The photocenter of the coma is located at the center of each panel (0, 0). Patches of bright and dark stains aligned in a certain direction are caused by background star tracklets. In Epoch $\sharp$2, central $\sim$3-by-3 pixels are saturated. 
\label{Fig03}}
\end{figure*}

$P_{\rm r}$ distributions across the coma (polarization maps) of K2 in the FILT\_691\_55 (Red domain) and FILT\_834\_48 (I-filter domain) are presented over the cometocentric distance $\rho$ in kilometers at three epochs of observations in Figure \ref{Fig03}. The uncertainties in the corresponding $P_{\rm r}$ ($\delta P_{\rm r}$) were calculated by error propagation and given in Appendix \ref{sec:app1-1} (Fig. \ref{Figap1-1}). The $\sim$3-by-3 central pixels at Epoch $\sharp$2 are saturated, causing difficulty in finding the true photocenter of the comet. Inaccurate alignment of the polarimetric images can lead to spurious features close to the photocenter \citep{Gray2024}. As such, instead of defining a center by eye, we used \texttt{DAOStarFinder}, a point-spread function (psf)-fitting photometry python subpackage of \texttt{photutils.psf}\footnote{\url{https://photutils.readthedocs.io/en/stable/psf.html}}, to determine the photocenter of the comet in each HWP angle strip. Images were searched for local density maxima with peaks greater than a threshold using a defined two-dimensional Gaussian kernel. We compared the outputs by changing Full Width at Half Maximum (FWHM) from 5 to 15 pixels and all test cases produced consistent pixel coordinates of the center mostly in sub-pixel accuracy. A (possibly) lower accuracy in estimating the central point compared to non-saturated cases and a non-uniform shape of the saturated area for each strip left a weird feature in the central $\sim$three by three-pixel area ($\sim$0.6\arcsec$^2$), which is marked as NaN in Figure \ref{Fig03} and excluded from further consideration.
As with the intensity distribution, K2 displays broadly homogeneous $P_{\rm r}$ distribution over its coma in both spectral domains throughout our observations, within the concentric coma region around the photocenter with a radius $\rho$ up to $\sim$2 $\times$ 10$^4$ km. The coma lacks any anisotropic polarimetric features, such as jets, swirls, or circumnuclear halo seen in the coma of several active comets (e.g., \citealt{Jones2000,Hadamcik2003a, Kiselev2020}). In this apparent constant $P_{\rm r}$ region, the null parameters $N_{Q}$ and $N_{U}$ (Figs. \ref{Figap1}$-$\ref{Figap3}) cluster around zero except for the background stars' streaks, validating the observed polarimetric features. K2 possessed negative $P_{\rm r}$ as expected based on its observation geometry of small $\alpha$ of $\lesssim$22\arcdeg\ (so-called `Negative polarization branch', NPB; \citealt{Kiselev2015} for a review).

\begin{figure*}[!t]
\centering
\includegraphics[width=0.9\textwidth]{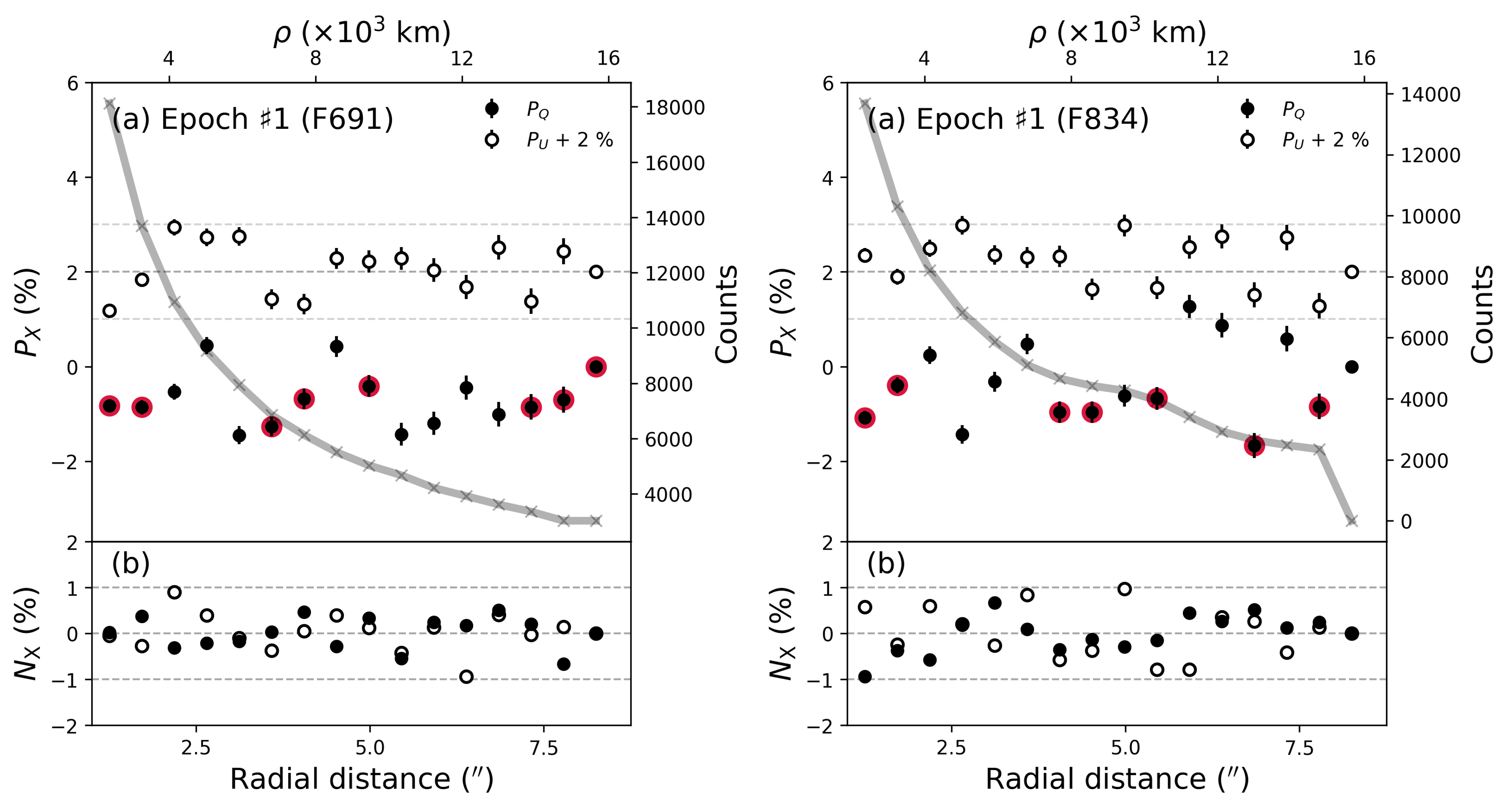}
\caption{
K2's Radial $P_{\rm r}$ distribution across a 4- to 34-pixel radius circle around the photocenter in the Red (left) and I-filter (right) domains at Epoch $\sharp$1. We derived local values within each annulus of two pixels in width. $P_U$ (offset $+$2 \% for clarity) and null parameters ($N_Q$ and $N_U$ in the same color scheme as $P_Q$ and $P_U$, respectively) are plotted together for validity check of the derived $P_Q$ values. Intensity profiles in counts are also given. $P_Q$ with red edges indicates data points selected according to the criteria mentioned in the main text. Due to overlapping with background stars, there are apparent fluctuations in the distributions of the parameters. 
\label{Fig04}}
\end{figure*}

\begin{figure*}[ht]
\centering
\includegraphics[width=0.9\textwidth]{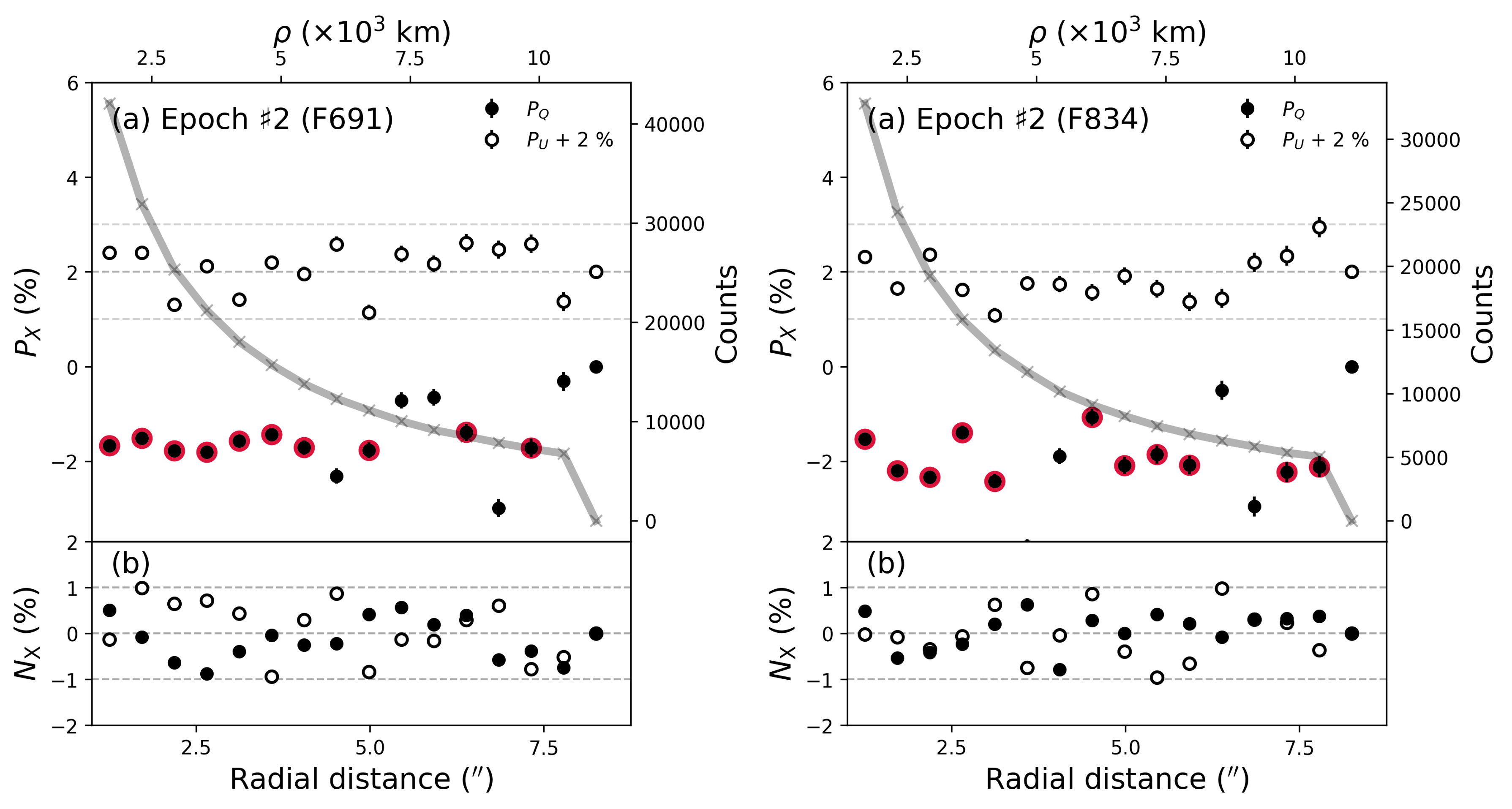}
\caption{Same as Figure \ref{Fig04} but for the Epoch $\sharp$2 data.
\label{Fig05}}
\end{figure*}

\begin{figure*}[ht]
\centering
\includegraphics[width=0.9\textwidth]{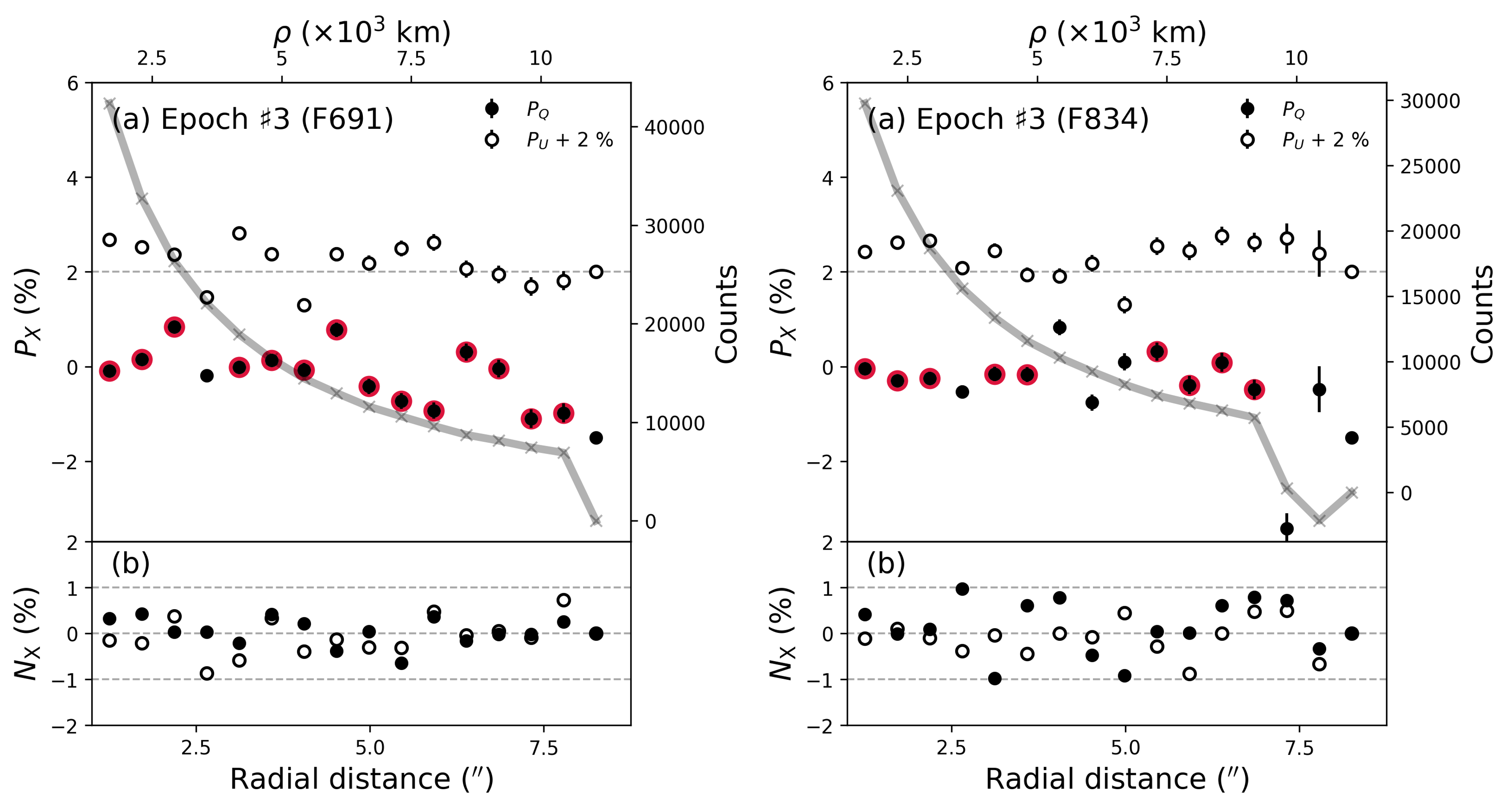}
\caption{Same as Figure \ref{Fig04} but for the Epoch $\sharp$3 data.
\label{Fig06}}
\end{figure*}

Given the lack of azimuthal anisotropy in $P_{\rm r}$, our next step is to examine the radial $P_{\rm r}$ distribution by measuring the average local $P_{\rm r}$ value. Local $P_{\rm r}$ values were measured every 2 pixels ($\sim$0.51\arcsec) in width increments over a circle with a radius of 4 to 34 pixels. The central parts ($<$4 pix) were omitted to prevent possible spurious measurements as a result of photocenter saturation. Figures \ref{Fig04}$-$\ref{Fig06} display the radial distribution of the weighted average and standard deviations of $P_{\rm r}$ (represented by $P_{Q}$), as well as the radial profiles of the $P_{U}$, their null counterparts $N_{Q}$ and $N_{U}$, and the intensity at each epoch. Dust coma intensity increased significantly between the first and second epochs. An annulus (local) intensity of K2 across the coma region analyzed ranges from inward to outward $\sim$68--15 (Red domain) and $\sim$53--11 (I-filter domain) for Epoch $\sharp$1, $\sim$114--23 (Red domain) and $\sim$87--17 (I-filter domain) for Epoch $\sharp$2, and $\sim$117--22 (Red domain) and $\sim$91--17 (I-filter domain) for Epoch $\sharp$3. Background star patches caused the measured values to fluctuate. The relatively center-dominated coma morphology led to a low S/N for the background coma. As a result, we selected valid data points only if they met all of the following conditions:
\begin{enumerate}
\item[$\bullet$] $|P_{U}|$ $<$ 1 \%~,
\item[$\bullet$] $|N_{Q}|$ $<$ 1 \% \& $|N_{U}|$ $<$ 1 \%~,
 \item[$\bullet$] Free from background star streaks.
\end{enumerate}
\noindent The selected data points are edged in red. 

The data distribution indicates that the $P_{\rm r}$ values (represented by $P_Q$ in Figs. \ref{Fig04}$-$\ref{Fig06}) in the inner and outer coma regions are constant within $\sim$2$\sigma$ at all epochs and thus, a single value would provide a moderate representation of the coma condition at a given epoch. We derived a weighted average of the red-edged data points at each epoch and the resulting values with their weighted standard errors over the considered range of radial distance in each filter ($P_{\rm r, 691}$ and $P_{\rm r, 834}$) are summarized in Table \ref{t02}. 

A variety of $P_{\rm r}$ radial gradient patterns have been observed on active comets to date, ranging from invariant to steeply changing. K2's nearly homogeneous $P_{\rm r}$ distribution within 2$\sigma$ over the coma appears consistent with several active comets whose $P_{\rm r}$ maps were constructed over similar radial distances in the NPB. For instance, long-period comet C/1990 K1 (Levy) has a nearly constant $P_{\rm r}$ distribution within the error bars over $\rho$ = 1,100$-$2,500 km at $\alpha$ $\sim$18$-$21\arcdeg\ (\citealt{Renard1992}, whose measurements are available from the PDS comet polarimetry catalog), although a different study on the same comet showed inhomogeneities in $P_{\rm r}$ across the coma at similar radial scales \citep{Hadamcik2003a}. A single-epoch data of C/2000 WM1 (LINEAR) at $\alpha$ = 12.6\arcdeg\ also showed a rather homogeneous coma in $P_{\rm r}$ up to $\rho$ $\sim$ 13,500 km \citep{Jockers2002}, yet caution is needed in light of their poor atmospheric seeing. Short-period comet 78P/Gehrels also exhibited a bland $P_{\rm r}$ distribution at $\alpha$ = 15.2\arcdeg, with the constant $P_{\rm r}$ from $\rho$ = 2,160$-$20,520 km \citep{Choudhury2014}.

\begin{deluxetable*}{c|cc|cc|c}[h]
\tabletypesize{\small} 
\tablewidth{0pt} 
\tablecaption{A summary of the polarimetric parameters of K2 at three epochs of VLT/FORS2 observations \label{t02}}
\tablehead{
\colhead{\multirow{2}{*}{Epoch}}  \vline&
\colhead{$\alpha$}  & 
\colhead{$\rho$}  \vline&
\colhead{$P_{\rm r, 691}$$^{(1)}$} &
\colhead{$P_{\rm r, 834}$$^{(1)}$}  \vline&
\colhead{$\nabla P_{\rm r}$$^{(2)}$}  \\
 & (\arcdeg) & (km) & (\%) &(\%) & (\% / 100 nm)
}
\startdata
1 & 15.9 & 1,600 -- 16,130 & $-$0.84 $\pm$ 0.07 & $-$0.89 $\pm$ 0.13 & $-$0.04 $\pm$ 0.11 \\
2 & 10.5 & 1,350 -- 11,440 & $-$1.69 $\pm$ 0.03 & $-$1.91 $\pm$ 0.12 & $-$0.15 $\pm$ 0.09 \\
3 & 20.0 & 1,340 -- 11,400 & $-$0.01 $\pm$ 0.14 & $-$0.16 $\pm$ 0.06 & $-$0.10 $\pm$ 0.11 
\enddata
\tablenotetext{1}{ Weighted average and weighted standard errors of the polarization degree of the coma averaged over the physical distance $\rho$ in the Red ($P_{\rm r, 691}$) and I-filter ($P_{\rm r, 834}$) domains. The data points used to derive the results are highlighted in Figures \ref{Fig04}$-$\ref{Fig06} in red based on the criteria described in the main text.}
\tablenotetext{2}{ Spectral gradient of polarization (or polarization color).} 
\end{deluxetable*}

\subsection{Polarization-phase curves} \label{sec:res2}

A directional dependence of cometary dust scattering results in the observed $P_{\rm r}$ varying with observing geometry and thus with phase angle $\alpha$, forming the so-called `polarization-phase curve' $P_{\rm r}(\alpha)$. Low reflectance with the irregular, agglomerate structure of cometary dust generally yields a bell-shaped $P_{\rm r}(\alpha)$ \citep{Mishchenko2010,Kiselev2015}; trends in the NPB at $\alpha$ $\lesssim$ 22\arcdeg\ of interest can be in particular used to characterize dust environments as it reflects a unique pattern of interference between scattered light waves determined by microscale physical and compositional properties of dust constituents \citep{Muinonen2015}. We analyzed the ensemble properties of the dust coma of K2 in comparison with those of other comets at similar $\alpha$ by using the NASA/PDS comet polarimetric archive \citep{Kiselev2017} accommodating $\sim$65 comets observed from 1940 to 2017 and the literature data published after 2017. We then fit the average coma trend in a spectral domain bracketed by the Red (620$-$730 nm) and I-filter (730$-$900 nm) domains with a conventional trigonometric function \citep{Lumme1993,Penttila2005}:
\begin{equation}
P_{{\rm r}, \lambda}(\alpha) = b \sin^{c_1}(\alpha) \cos^{c_2}\Big(\frac{\alpha}{2}\Big) \sin(\alpha - \alpha_0)
~,
\label{eq:eq4}
\end{equation}
\noindent where $b$, $c_1$, $c_2$, and an inversion angle $\alpha_0$ (the angle at which $P_{\rm r}$ changes its sign and thus $P_{\rm r}$ = 0 \%) are wavelength-dependent parameters. Best-fit parameters are those with minimum $\chi^2$ for the curve.

Data sets were selected for analysis based on several criteria. Firstly, we chose either narrowband dust coma measurements or measurements for which flux from gas emissions (e.g., C$_{2}$, NH$_{3}$, and CN-red; \citealt{Kwon2018}) has been carefully examined to demonstrate their negligible contribution to the total observed signals. This criterion helps minimize potential contamination impacts of photo-dissociated gas molecules on determining the inherent dust continuum polarization. We excluded comet observations conducted at $r_{\rm H}$ $\lesssim$ 2 au if the above information was not available.
Secondly, when multiple aperture $P_{\rm r}$ values are available for a single comet at the given $\alpha$, only measurements with apertures similar to those of the K2 ($\rho$ $\sim$ 1,500$-$15,000 km) have been selected for this study. In cases where a comet's $P_{\rm r}$ is nearly invariant with aperture size, we selected data in the middle of the aperture ranges considered in the reference. 
Finally, C/1995 O1 (Hale-Bopp) and 2I/Borisov were not used for fitting because of their large $P_{\rm r}$ deviations from those of typical comets in the given phases (e.g., \citealt{Hadamcik2003b,Bagnulo2021}), let alone measurements of K2 (Table \ref{t02}) and comet nuclei.

\begin{figure}[!b]
\centering
\includegraphics[width=0.87\textwidth]{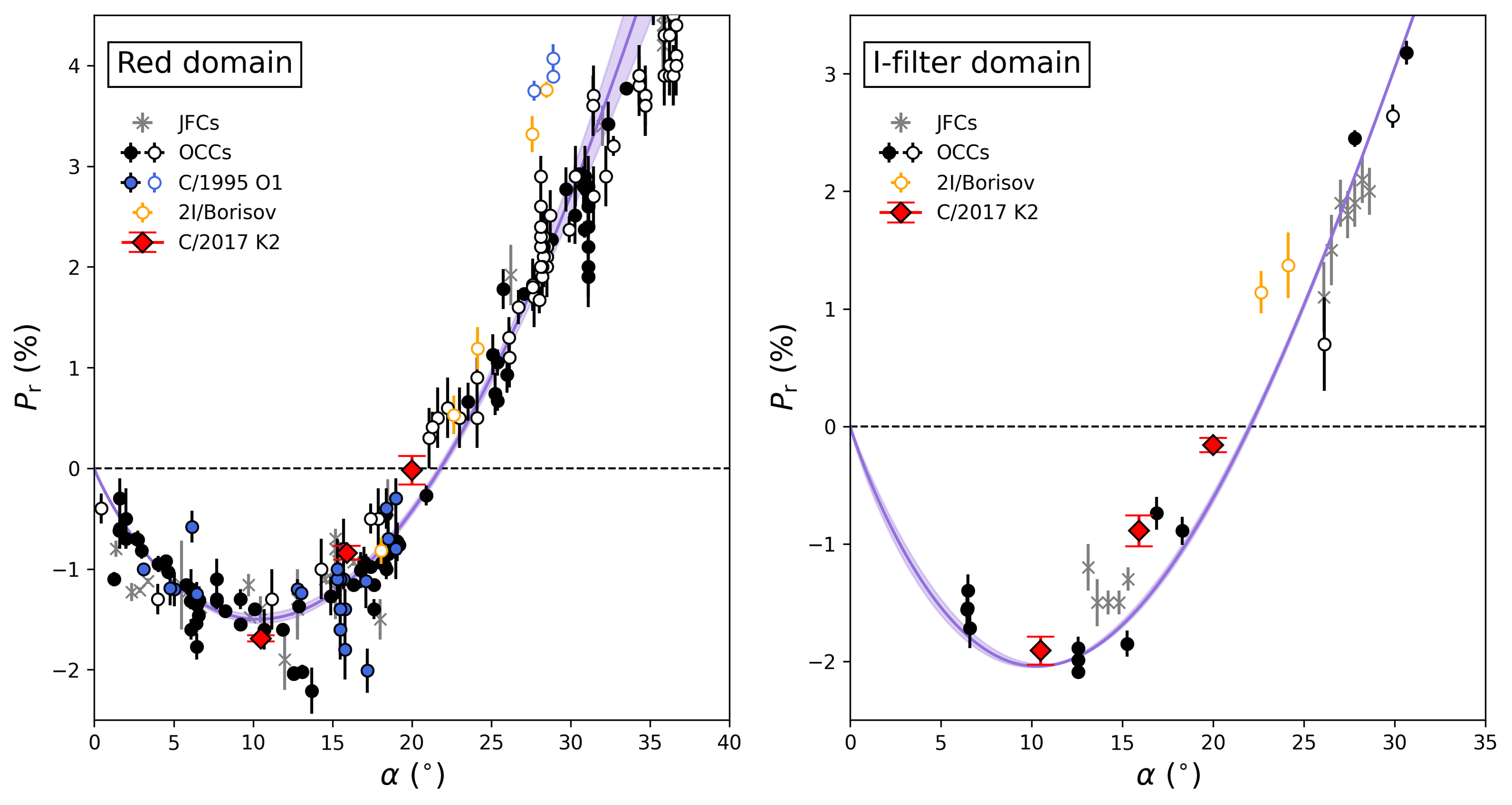}
\caption{Polarization-phase curve of cometary dust in the Red domain (620$-$730 nm, left panel) and the I-filter domain (730$-$900 nm, right panel) quoted from the NASA/PDS comet polarimetric archive (\citealt{Kiselev2017} and references therein) and the literature published afterward: 67P/Churyumov-Gerasimenko \citep{Rosenbush2017,Kwon2022} which were lumped into the category of JFCs and 2I/Borisov \citep{Bagnulo2021}. Preperihelion (inbound) and postperihelion (outbound) data were distinguished with filled and open symbols, respectively. The solid curve and its shade area indicate the fitted average trend (Eq. \ref{eq:eq4}) and its 1$\sigma$ error, respectively. The number of data points used for fitting is 29 for JFCs and 181 (=96 inbound + 85 outbound) for OCCs in the Red domain and 17 for JFCs and 21 (=19 inbound + 2 outbound) for OCCs in the I-filter domain.
\label{Fig07}}
\end{figure}

A collective polarization-phase curve of coma dust in the Red and I-filter domains is shown in Figure \ref{Fig07}, with data for long-period comets observed in inbound (filled circles) versus outbound (open circles) legs distinguished, as well as Jupiter-family comets (JFCs) versus Oort-cloud comets (OCCs, including Halley type comets). The number of data points selected for the JFCs is insufficient to examine their separate trends preperihelion versus postperihelion.
The best-fit parameters which depict the ensemble average $P_{\rm r}(\alpha)$ of dust comae are $b$ = 0.36 $\pm$ 0.03, $c_{1}$ = 0.90 $\pm$ 0.03, $c_{2}$ = 2.13 $\pm$ 3.17, and $\alpha_{0}$ =21.80\arcdeg\ $\pm$ 0.18\arcdeg\ in the Red domain; and $b$ = 0.50 $\pm$ 0.04, $c_{1}$ = 0.92 $\pm$ 0.05, $c_{2}$ = 5.40 $\pm$ 0.79, and $\alpha_0$ = 22.03\arcdeg\ $\pm$ 0.12\arcdeg\ in the I-filter domain. 
Differentiating the average curve and finding a point on the curve whose derivative becomes zero, $P_{\rm r}(\alpha)$ in the NPB has a minimum $P_{\rm r}$ ($P_{\rm min}$) of $-$1.50$^{+0.05}_{-0.04}$ \% at $\alpha_{\rm min}$ = 10.35\arcdeg$^{+0.11\arcdeg}_{-0.12\arcdeg}$ in the Red domain and of $-$2.04$^{+0.02}_{-0.01}$ \% at $\alpha_{\rm min}$ = 10.32\arcdeg$^{+0.31\arcdeg}_{-0.32\arcdeg}$ in the I-filter domain.

K2's $P_{\rm r}$ values at all epochs ($\alpha$ = 15.9\arcdeg, Epoch $\sharp$1; $\alpha$ = 10.5\arcdeg, Epoch $\sharp$2; and $\alpha$ = 20.0\arcdeg, Epoch $\sharp$3) are broadly consistent with those of other OCCs at similar observing geometry, though a slight $P_{\rm r}$ excess on K2 at Epoch $\sharp$3 more resembles the case of C/1995 O1 (Hale-Bopp) and 2I/Borisov. There was no statistically significant difference in $P_{\rm r}(\alpha)$ between JFCs and OCCs. Postperihelion observations of OCCs in the Red domain have a slightly shallower depth in the NPB within 2$\sigma$ and a smaller $\alpha_0$ ($P_{\rm min}$ = $-$1.56$^{+0.17}_{-0.04}$ \%\ and $\alpha_0$ = 19.8\arcdeg\ $\pm$ 0.3\arcdeg) than preperihelion ($P_{\rm min}$ = $-$1.85$^{+0.03}_{-0.02}$ \%\ and $\alpha_0$ = 22.6\arcdeg\ $\pm$ 0.1\arcdeg). The I-filter domain shows no corresponding differences as there are few measurements available and the data points are too scattered to draw meaningful conclusions.
\\

\subsection{Spectral gradient of polarization} \label{sec:res3}

Observations in spectropolarimetry or (quasi-)simultaneous multiband aperture polarimetry enable us to estimate a spectral gradient of the polarization $P_{\rm r}(\lambda)$, or the so-called `Polarization color (PC)' in \% per 100 nm units using the following equation:
\begin{equation}
{\rm PC} \equiv  \nabla P_{\rm r} = \frac{P_{{\rm r}, \lambda_2} - P_{{\rm r}, \lambda_1}}{\lambda_2 - \lambda_1}~,
\label{eq:eq5}
\end{equation}
\noindent where $P_{{\rm r}, \lambda_W}$ indicates $P_{\rm r}$ in \% measured at a wavelength $\lambda_{\rm W}$ in nanometers ($\lambda_2$ $>$ $\lambda_1$). Traditionally, a positive PC is labeled red, while a negative PC is labeled blue. As K2's $P_{\rm r}$ values were measured on the same night in the Red and I-filter domains, we could derive its dust coma PC at each epoch, assuming the rotation-induced variation in  $P_{\rm r}$ in the coma is negligible.

\begin{figure}[!b]
\centering
\includegraphics[width=0.475\textwidth]{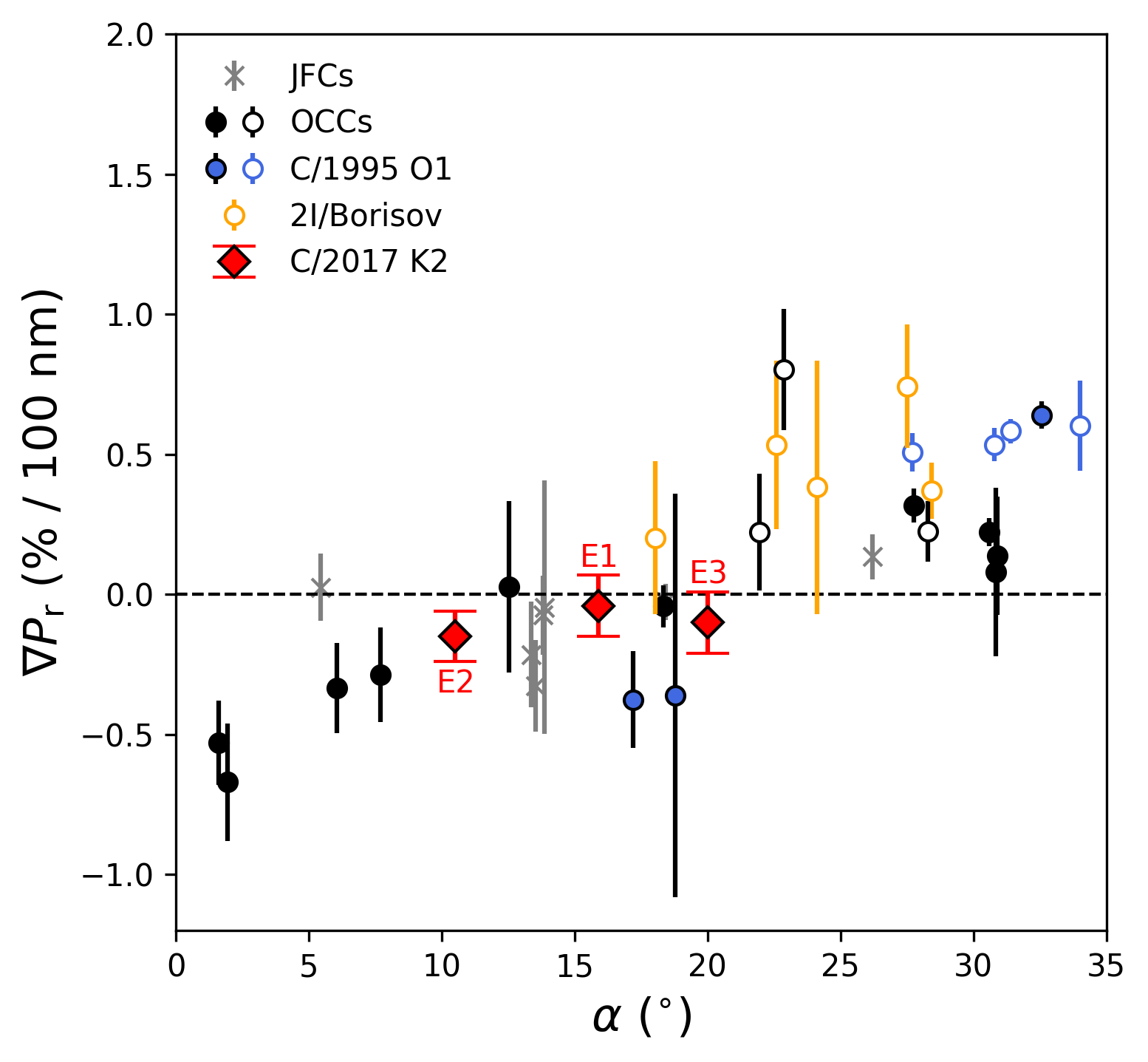}
\caption{Spectral gradient of polarization or polarization color (PC) of cometary dust as a function of phase angle $\alpha$. Preperihelion and postperihelion measurements are distinguished by filled and open circles, respectively. Epochs $\sharp$1, $\sharp$2, and $\sharp$3 are denoted by E1, E2, and E3, respectively.
\label{Fig09}}
\end{figure}

Figure \ref{Fig09} shows the PC of cometary dust with $\alpha$, including our K2 results (Table \ref{t02}). Based on the same criteria applied to the polarization-phase curve analysis (Sect. \ref{sec:res2}), we selected $P_{\rm r}$ measurements for comets whose multiband $P_{\rm r}$ values have been measured simultaneously or at least on the same night and calculated their PC. Considering that the PC of cometary dust is in general linear at optical wavelengths (between $\sim$400 and $\sim$1,000 nm, equivalent to BVRI-filter domains; \citealt{Kiselev2015} for a review), we directly compared K2's PC measured over R$-$I with the PCs of other comets derived anywhere in the optical spectral range (e.g., over V$-$R and V$-$I). The measurements used here came from the following references: 1P/Halley, 17P/Holmes, C/1995 O1 (Hale-Bopp), C/2009 P1 (Garradd), and C/2000 WM1 (LINEAR) from the NASA/PDS archive \citep{Kiselev2017}, 2I/Borisov \citep{Bagnulo2021}, and 67P/Churyumov-Gerasimenko \citep{Kwon2022}. 

In Figure \ref{Fig09}, the PC of cometary dust overall increases linearly with $\alpha$, leading to an increase in the absolute value of PC as $\alpha$ shifts away from $\alpha_{0}$, as already shown in previous studies of cometary dust (e.g., \citealt{Kolokolova2004,Kiselev2015,Kwon2022}) and in regolith particles of asteroids (e.g., \citealt{Cellino2015,Kwon2023b}). The PC of K2 at all epochs remains well within the scatter of other comets observed under similar conditions. The broad consistency of PC indicates the dust particles responsible for K2's inner coma dust explored in polarimetry are likely to have ensemble properties similar to those of active comets, or at least similar across optical domains.
\\

\subsection{H$\alpha$ emission} \label{sec:res4}

In conjunction with dust continuum polarimetry, a narrowband filter H\_Alpha$+$83 was used to observe three epochs of the H$\alpha$ emission line peaking at 656.3 nm (Table \ref{t01}). As H$\alpha$ emission has shown its relevance to emissions from water molecules in the comae of several active comets (e.g., \citealt{Combi1999,Swamy2010,Bodewits2019}), monitoring K2's H$\alpha$ emission during its trip to perihelion would provide insight into the extent to which water-ice sublimation contributes to cometary activity.
We made seven 30-sec-exposure H$\alpha$ images at each epoch, which were coadded into a composite image; yet, the morphology of the observed H$\alpha$ coma appeared nearly identical to that of the adjacent continuum (FILT\_691\_55) image. Red-domain dust brightness always significantly outweighed H$\alpha$. In the scaled and subtracted image (=scaled H$\alpha$ brightness $-$ dust brightness), the remaining H$\alpha$ feature was too faint to be measured, possibly due in part to insufficient sensitivity. A negligible contribution from K2's H$\alpha$ emission to the observed light is compatible with its MUSE spectrum (over 420$-$950 nm) obtained at 2.53 au (the same date as Epoch $\sharp$3 of our VLT observations), which lacks emission features around the wavelength expected \citep{Kwon2023a}. We will thus no longer discuss the results of H$\alpha$ photometry in the following sections.
\\

\section{Discussion} \label{sec:dis}

The discontinuous brightening event observed at $\sim$2.9 au indicates a possible change in the activation state of K2. This section examines the secular evolution of its coma activity (Sect. \ref{sec:dis1}) in light of the dust properties that dominate the coma signals (Sect. \ref{sec:dis2}), aided by numerical modeling of dust light scattering. All short- and long-term results are combined to discuss the dust environment of the nucleus of K2 in a broader context (Sect. \ref{sec:dis3}).
\\

\subsection{Accelerating activation of K2 around 2.9 au} \label{sec:dis1}

\begin{figure}[ht]
\centering
\includegraphics[width=0.95\textwidth]{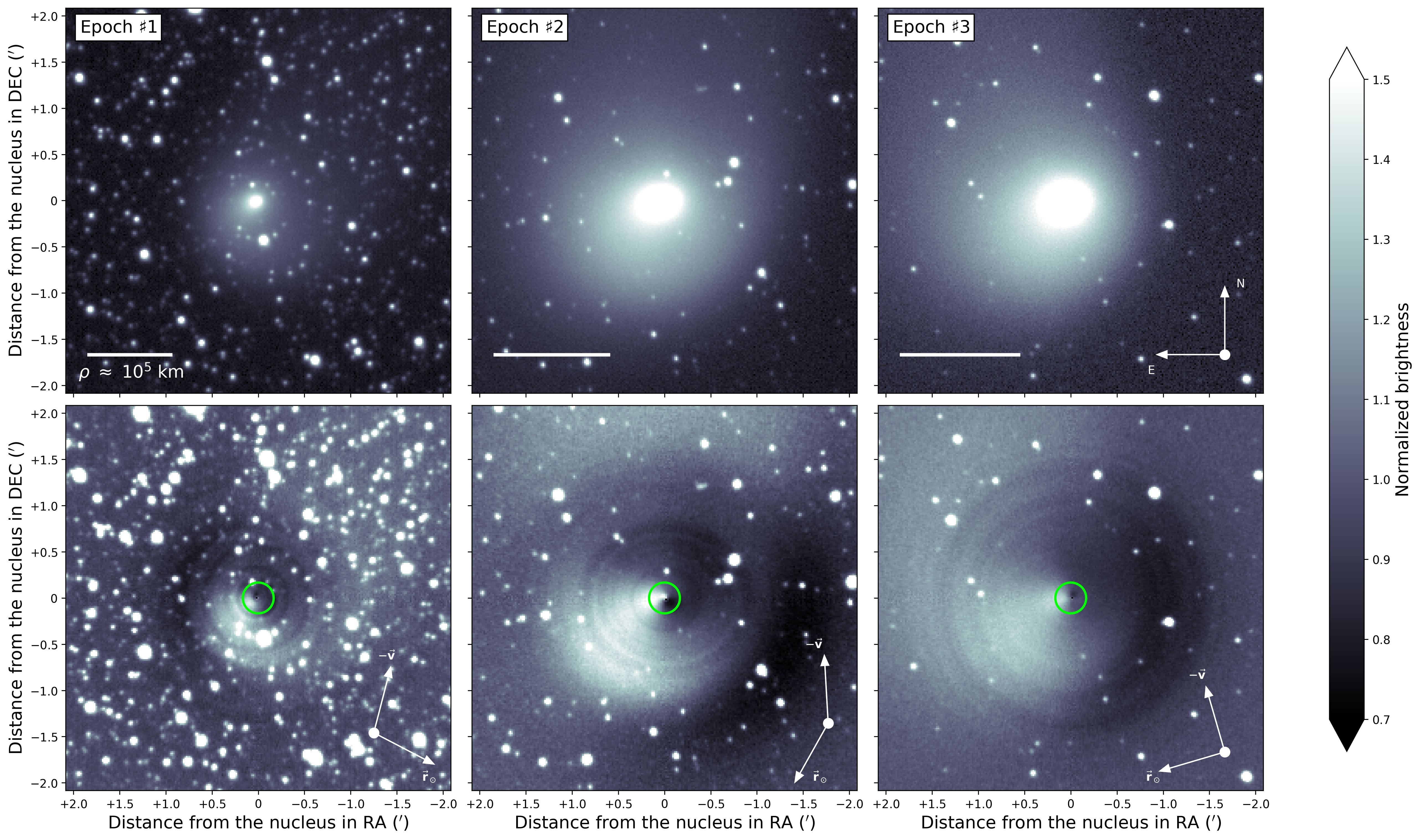}
\caption{ZTF zr-band intensity image of K2 (upper row) and its enhanced version divided by azimuthal median (lower row) taken at or near the time of VLT polarimetric observations: on UT 2022 May 11.4 for Epoch $\sharp$1, UT 2022 June 26.4 for Epoch $\sharp$2, and UT 2022 July 19.3 for Epoch $\sharp$3. All images were normalized on the same brightness scale and aligned with the North on top and the East on the left. Anti-sunward ($\vec{\bf{r}}_\sun$) and negative velocity ($-\vec{\bf{v}}$) vectors at each epoch are given. Green circles in the center encircle the outer boundary of the polarimetric aperture.  
\label{Fig10}}
\end{figure}

The heliocentric brightening trend of K2 preperihelion shown in Figure \ref{Fig01} indicates that at around 2.9 au, the coma photometric parameters changed substantially, accelerating brightness and dust mass loss. Figure \ref{Fig10} presents the ZTF zr-band image of K2 and its enhanced version on the upper and lower rows, respectively. Three images covering before, around, and after the discontinuous activation were selected on the same date or the closest available date to our VLT polarimetric observations. By dividing intensity images by azimuthal median, we searched for anisotropic coma features. Since the median is by nature less affected by outliers than the average, and our $P_{\rm r}$ are nearly consistent in a radial direction (Figs. \ref{Fig04}$-$\ref{Fig06}), we implemented this method to detect azimuthal changes embedded in an outer coma envelope (e.g., \citealt{Samarasinha2014}). The green guiding circles in the center encompass the outer edge of the polarimetric aperture, covering the inner coma part. The photometric parameters in Figure \ref{Fig01} derived with $\rho$ = 20,000 km on the intensity images represent the large-scale coma information collected over an area two orders of magnitude larger than the polarimetric aperture. It is interesting to note that the outer coma component heads in a negative velocity direction $-\vec{\bf{v}}$ while the inner coma component is anti-sunward $\vec{\bf{r}}_\sun$ (upper row in Fig. \ref{Fig10}) and such misalignment becomes more apparent as K2 progressed from Epoch 1 to Epoch 2, likely associated with the discontinuous activation at $\sim$2.9 au. An anti-sunward plume structure embedded within the outermost coma envelope (lower row in Fig. \ref{Fig10}) became more intense during this period, as shown by \citet{Kwon2023a}.

\begin{figure}[ht]
\centering
\includegraphics[width=0.3\textwidth]{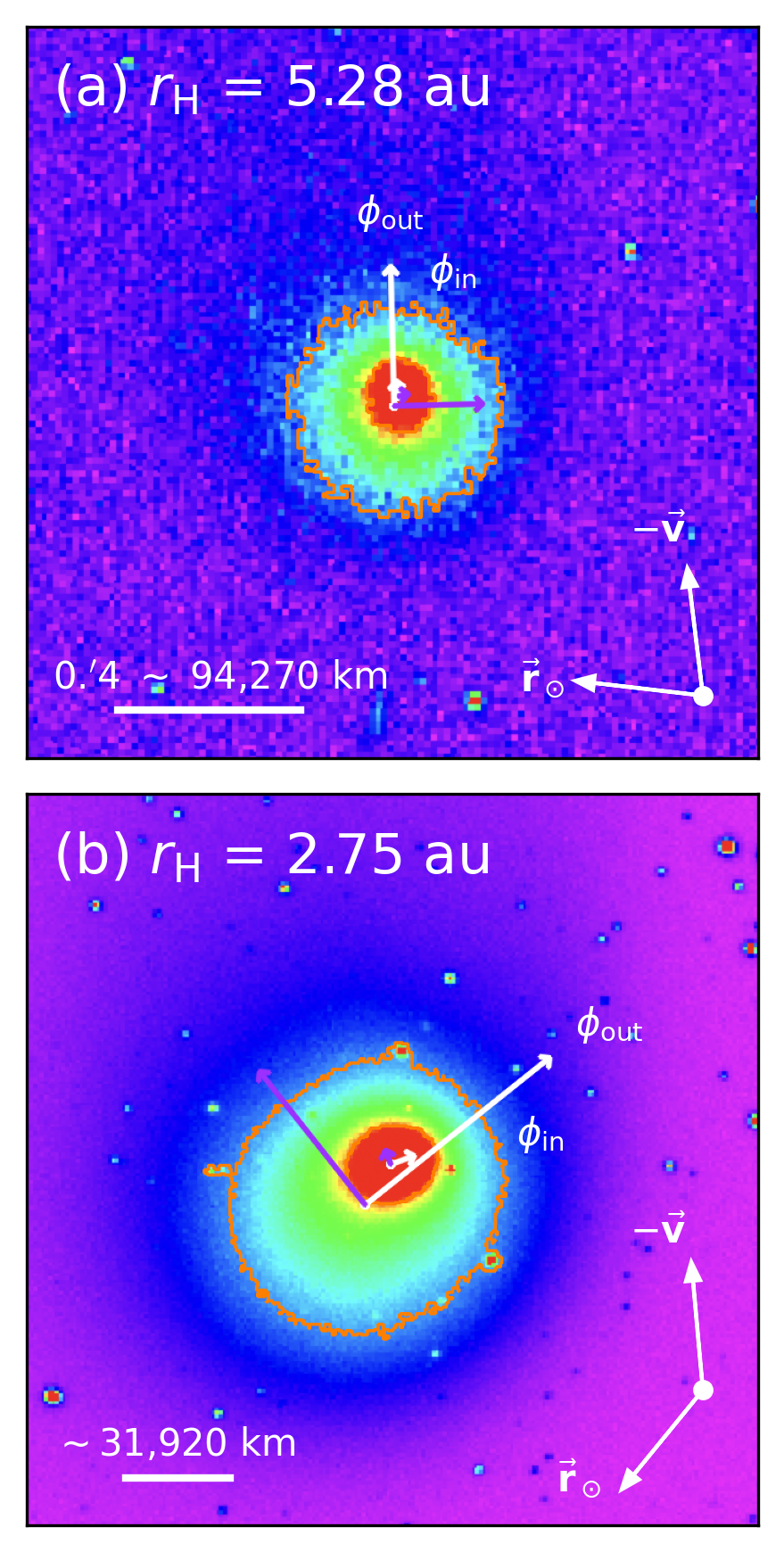}
\caption{PCA example. Primary (PC1) and secondary (PC2) eigenvectors are shown as white and purple arrows, respectively. $\phi_{\rm out}$ and $\phi_{\rm in}$ denote a primary axis' position angle of the outer and inner coma components, respectively. Two epochs were arbitrarily chosen as examples in which the inner and outer coma components presented similar or different angle alignments. \textcolor{red}{Scale bars at the bottom show the projected physical distance from the comet center in kilometers equivalent to 0.4 arcminutes at each epoch.}
\label{Fig11}}
\end{figure}

Motivated by a possible association between the anti-sunward plume launch near the nucleus and the distinct directionality development of the inner coma, we examined how the overall coma morphology evolved over heliocentric distances.
To this end, we applied the principal component analysis (PCA) in \texttt{PCACompute2} of \texttt{OpenCV} python package, treating a coma as a two-dimensional oval-shaped projection whose morphology is well depicted by two principal components (PC1 and PC2). PC1 and PC2 indicate the semi-major and semi-minor axes of the coma extension, respectively. Position angles ($\phi$) for the primary and secondary were measured counterclockwise from the north. Using the applied technique, we identified a coma as an independent component and measured its eigenvectors and eigenvalues only when the count exceeded $\sim$3$\sigma$ of the ambient background and the extension exceeded three times FWHM at the given date (\texttt{SEEING} in header keyword). Figure \ref{Fig11} shows two examples, one without segregation and the other with a well-developed inner coma component standing out from the outer coma component. The coma morphology was considered concentric if the primary axis (PC1 eigenvector) has $\phi$ = 0\arcdeg\ or 90\arcdeg\ (or 180\arcdeg\ or 270\arcdeg, respectively).

\begin{figure}[ht]
\centering
\includegraphics[width=0.7\textwidth]{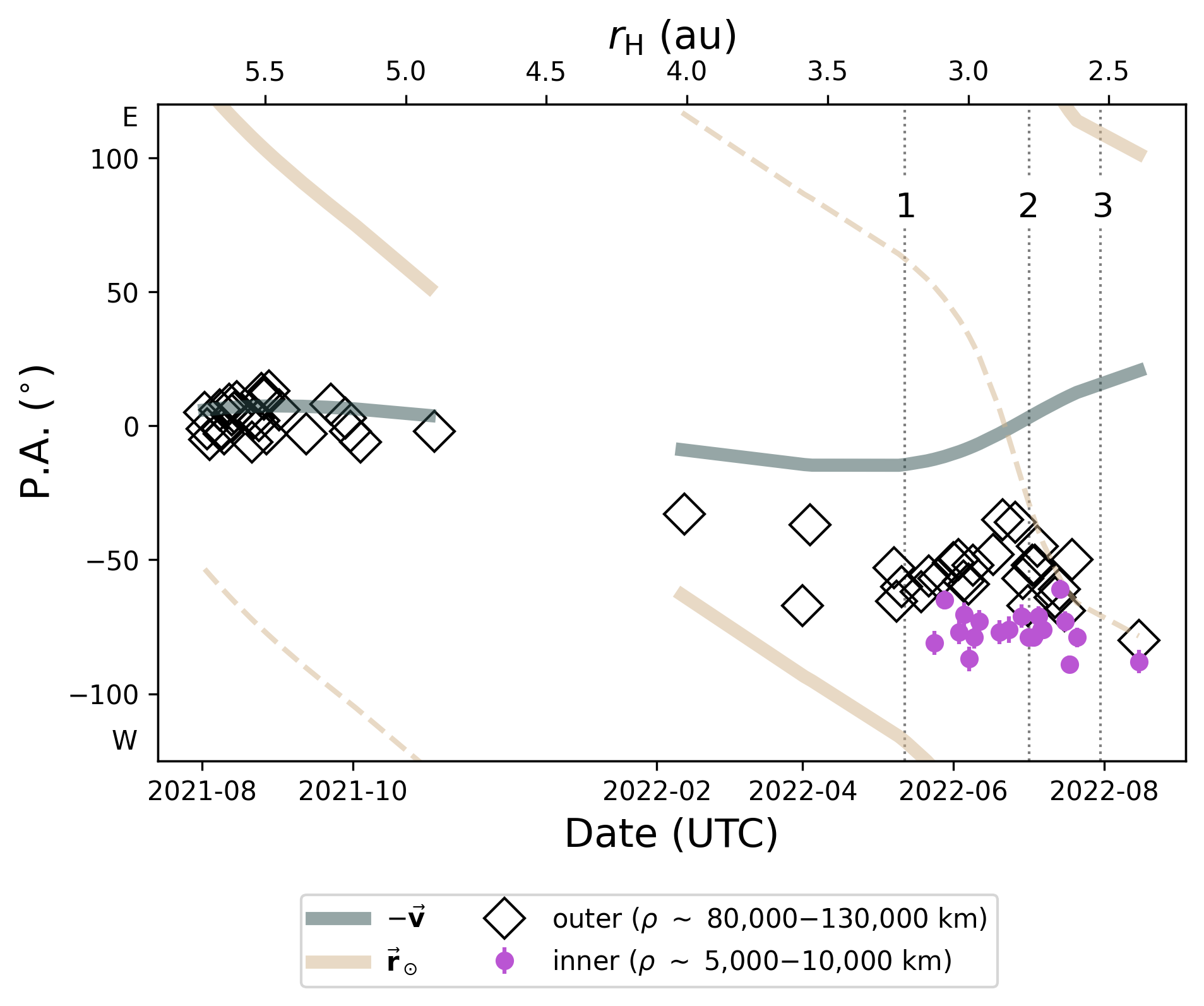}
\caption{Heliocentric evolution of the semi-major axis' position angle (the PC1 eigenvector) of K2 coma. The solid lines indicate the anti-sunward $\vec{\bf{r}}_\sun$ (brown) and negative velocity $-\vec{\bf{v}}$ vectors (green), while the dashed lines in the same colors show 180\arcdeg\ apart directions. We plotted inner coma components (filled circles) only when they were aligned differently from the outer coma components (open diamonds), omitting circles well inside the diamonds for simplicity. Three epochs of our VLT polarimetric observations are marked with dotted vertical lines. 
\label{Fig12}}
\end{figure}

Figure \ref{Fig12} shows the secular evolution of the position angles of the coma components of K2. The coma became wide enough to measure when the comet entered within $r_{\rm H}$ $\sim$ 6 au. There was no discernible separation in the coma's alignment until $\sim$3 au (around Epoch $\sharp$1); the dust tail broadly followed the direction of K2's negative velocity. As K2 crossed the outer boundary of the water ice line ($\lesssim$3 au, thus between Epoch $\sharp$1 and Epoch $\sharp$2), the inner coma component gradually aligned with the comet's anti-sunward direction. In later phases (Epoch $\sharp$2 and Epoch $\sharp$3), the dust tail oriented approximately anti-sunward, including the inner coma component. Such a similar timeline in discontinuous evolution of the morphology and brightening of K2's coma provides supportive evidence that the comet was experiencing an accelerating change at $\sim$2.9 au.
Furthermore, the different heliocentric variations of the inner and outer coma components (Figs. \ref{Fig10} and \ref{Fig12}) underline the necessity to be cautious in using photometric and polarimetric parameters together to interpret the coma environment: signals of the former mostly originated from the outer coma and those of the latter more represent the inner state. The two observations cover different scales of the coma region and thus likely highlight different dust populations (e.g., \citealt{Marschall2020b}).  
\\

\subsection{Secular evolution of coma dust of K2} \label{sec:dis2}

Observed temporal evolution of K2 activity (Sect. \ref{sec:dis1}) suggests a change in scattering characteristics of the coma components. In this subsection, we attempt to constrain the dust properties dominating coma signals around each epoch of our polarimetric observations.

\subsubsection{Dust particles compatible with the observed activity evolution of K2} \label{sec:dis2-1}

Over the past decades, remote and in-situ missions have shown that cometary dust has broadly similar microscale profiles irrespective of dynamical groups; it consists primarily of absorbing carbonaceous materials (amorphous carbon and organics), Mg-rich silicates, Fe-sulfides, and other trace minerals aggregated into a hierarchical structure resulting in low albedo of a few percent (\citealt{Levassuer-Regourd2018} for a review). In contrast to the discontinuous activation of the large-scale coma at $\sim$2.9 au (Fig. \ref{Fig01} and Sect. \ref{sec:dis1}), polarimetric behaviors on the comet that were primarily influenced by scattered light from the inner portion of the coma show no significant deviation from the average trends of active comets at similar phases (Sect. \ref{sec:res1}$-$\ref{sec:res3}). Given the nature of polarimetry, K2's not-unusual polarimetric response suggests that the microscale properties of the dust population dominating the inner coma signal throughout the observations be fairly intact or do not significantly deviate from the expected trend.

Several dust properties can be excluded from consideration based on the observed polarimetric behaviors over the inner coma (an order of $\rho$ $\sim$ 10$^3$ km). Firstly, a large amount of Rayleigh-like small-end Mie-regime dust particles (size parameter X = 2$\pi$$a$ / $\lambda$ $\lesssim$ 1, where $a$ is the dust radius and $\lambda$ is the wavelength of observation, which corresponds to an order of $\sim$0.1$-$1 $\mu$m in the optical) in the inner coma would be a less appealing option since they can suppress NPB development. As an NPB around the backscattering region results from interference effects between the multiple-scattered light (coherent backscattering; \citealt{Muinonen2015} for a review), insignificant electromagnetic interactions between such small dust can suppress NPB developments. An extreme case would be the absence of the NPB of cometary comae having abundant Rayleigh particles (X $\ll$ 1), as seen in C/1995 O1 (Hale-Bopp) at infrared wavelengths \citep{Petrova2001,Kolokolova2007,Lasue2009,Kiselev2015}. Light scattering-wise, a coma dominated with highly porous dust aggregates (e.g., fluffy agglomerates having porosity of $\gtrsim$99 \%; \citealt{Guttler2019}) behaves similarly to a single sub-micron dust constituent \citep{Kolokolova2011}. Secondly, dust much larger than wavelengths (X $\gg$ 10, often an order of $\gtrsim$100 $\mu$m) can also shallow the NPB depending on the refractive index of the scattering material and thus less likely to predominate the signal of K2. Suppose dust particles in this scattering regime have both large real and imaginary parts of the refractive index, akin to dark, consolidated dust chunks often seen in comae of Jupiter-family comets \citep{Hadamcik2009}. Their large refractive indices favor a single-scattering environment restricted to the surface layer, minimizing second-order scattering effects between dust constituents \citep{Petrova2001,Hadamcik2009,Kwon2018}. Such dust could also contain water ice but to a limited extent (refractory fraction of $\gtrsim$1 \%; \citealt{Mukai1986}), so that the resulting optical properties resemble those of refractory materials. (By comparison, if significant amounts of ice are present in dust, as in the coma of 103P/Hartley 2 \citep{A'hearn2011}, optical properties follow those of ice.) 
Thirdly, in the same large particle regime, the significant contribution of dust with a relatively small $m_{\rm i}$ (transparent) to the total brightness, either as silicates or water ice (e.g., Fig. 3 in \citep{KolokolovaJoc1997}), is also incompatible with the observations. Multiple scattering of incident light among dust constituents would be enhanced by a smaller imaginary part of the refractive index ($m_{\rm i}$ $\lesssim$ 0.01) than typical cometary dust ($m_{\rm i}$ $\sim$ 0.1; e.g., \citealt{Moreno2018}), resulting in impaired alignment of scattered light (depolarization) and thus shallow the NPB. Little quantitative information is currently available regarding the threshold size and porosity of aggregated particles that define the scattering regimes of NPB enhanced by coherent backscattering versus NPB suppressed by depolarization.

\begin{figure}[ht]
\centering
\includegraphics[width=0.7\textwidth]{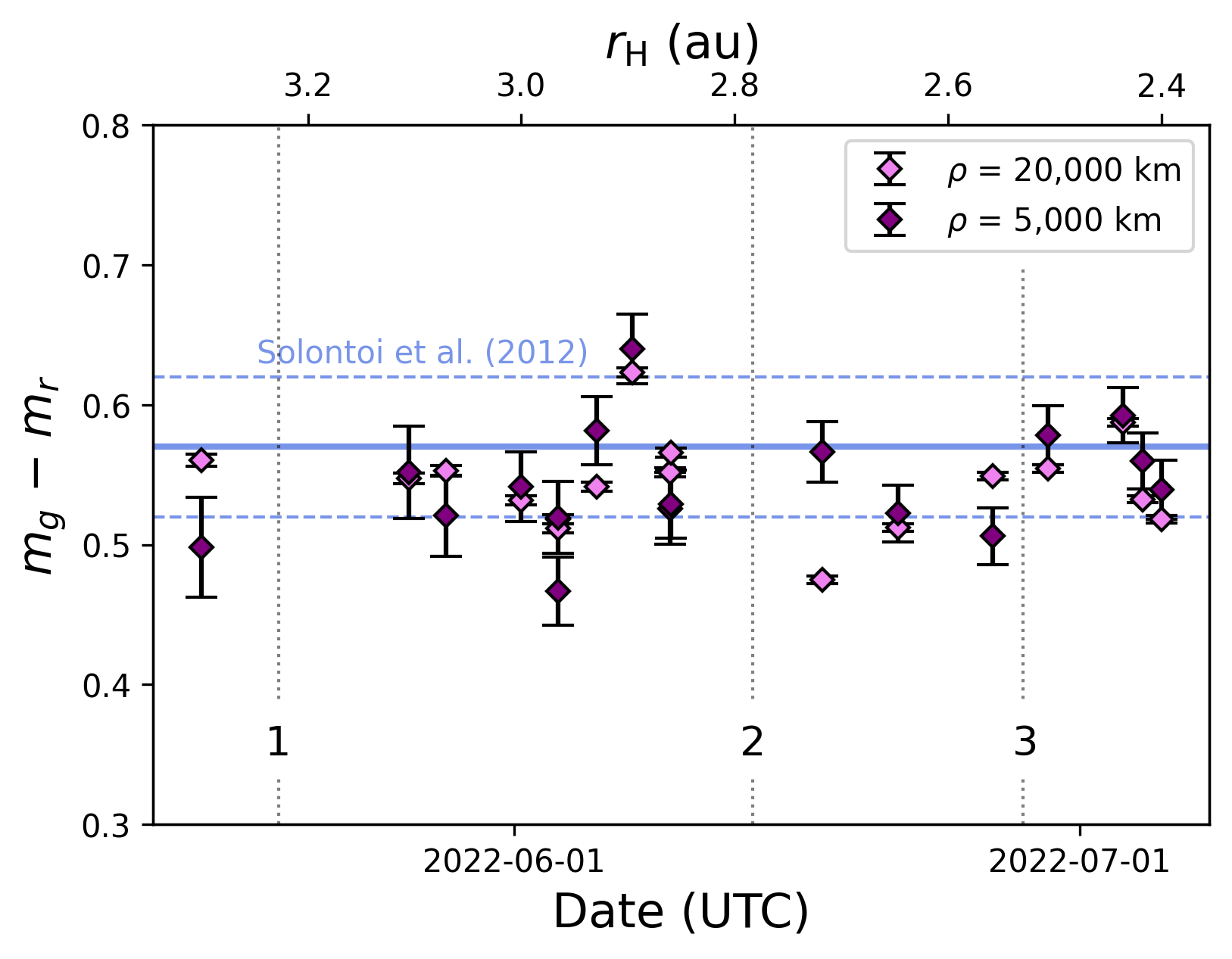}
\caption{Secular evolution of the {\sl{g}$-$$r$} color of K2 preperihelion. Colors were measured with two different aperture radii $\rho$, each equivalent to the coma regions covered by photometry (pink diamonds) and photometry (purple diamonds). Three numbered vertical lines denote the epochs of VLT polarimetric observations. Blue central solid and its two dashed horizontal lines indicate the average {\sl{g}$-$$r$} dust color of 31 active solar system comets and its standard deviations observed by the Sloan Digital Sky Survey \citep{Solontoi2012}.
\label{Fig13}}
\end{figure}

Dust color can provide additional constraints on dust in showing how refractive indices and size parameters vary with wavelength \citep{Hadamcik2009}. With the ZTF archive's zg- (effective wavelength 474.6 nm and FWHM 131.7 nm) and zr-band images of K2 taken on the same night, we conducted aperture photometry with two different aperture sizes of the coma (using Eq. \ref{eq:eq0} with $\rho$ of 20,000 km and 5,000 km, approximating radial scales of photometry in Fig. \ref{Fig01} and polarimetry in Fig. \ref{Fig03}, respectively), considering the heterogeneous coma condition of K2 preperihelion (Figs. \ref{Fig10}$-$\ref{Fig12}). The zr-band apparent magnitudes ($m_{r}$) were subtracted from the zg-band magnitudes ($m_{\sl g}$) to determine dust color. A summary of the heliocentric evolution of the {\sl{g}$-$$r$} color of K2 preperihelion is shown in Figure \ref{Fig13} on the two radial scales. The heliocentric distances of our VLT observations are marked, along with the average {\sl{g}$-$$r$} color and 1$\sigma$ deviation of 31 active solar system comets (0.57 $\pm$ 0.05; \citealt{Solontoi2012}).

Large fluctuations in color seen both in the inner and outer coma regions of K2 around 2.9 au (Fig. \ref{Fig13}) again support a discontinuous change in the comet's preperihelion activity. Our polarimetric observations do not directly cover this transitional period and are instead located when the coma colors fell into the 1$\sigma$ distribution. This is in line with the K2's polarimetric behavior in a typical range of active comets. The coma color decreased rapidly (bluing) right outside $\sim$2.9 au. This condition may be caused by abundant small dust particles or a large amount of water ice (similar to the dust properties in the first or third cases discussed above, respectively) in the coma. Immediately following this, the coma displayed a significantly redder color than the average. The temporal reddening in the coma, especially in the inner coma, may be caused by absorbing large dust chunks moving at low outward speed in the coma \citep{Hadamcik2009,Kwon2022}, where high-mobility dust particles either in small sizes or with high porosity were already swept by the enhanced water-ice activity around this term with high outward velocity. Two opposite scenarios participating in the observed color variation (small/transparent dust for the bluing and large/absorbing dust for the reddening) at $\sim$ 2.9 au would imply that the dust particles ejected around the term may represent a combination of a variety of properties. This idea seems compatible with the suggestion made by \citet{Fulle2022} that the relatively large size of K2's nucleus allows ejecting dust populations with a wide range of mobility distributions that can be parameterized by the inverse multiplication of $\rho$$r$ ($\rho$ is the mass density and $r$ is the size of the dust).

Gas emissions can contribute to the bluing of the outer coma at a later phase: a larger fraction of C$_{2}$ Swan band emissions flux to the total zg-band signal would increase as the comet drifted inward through Epoch $\sharp$2. Near the inner boundary of the water sublimation boundary (on the same date as Epoch $\sharp$3), K2 showed significant flux of the C$_2$(0,0) Swan band (9.4 $\pm$ 1.0) $\times$ 10$^{-12}$ erg cm$^{-2}$ s$^{-1}$ over the same area ($\rho$ = 20,000 km), which was negligible within the inner coma \citep{Kwon2023a}.

The secular evolution of K2's {\sl{g}$-$$r$} color, which shows a change within a broadly typical coma range of active solar system comets \citep{Solontoi2012}, was also compared to 26 active long-period comets (LPCs), 18 of which have perihelion distances that exceed Jupiter's semimajor axis of 5.2 au \citep{Jewitt2015}. This comparison might provide a hint about potential differences between K2's dust and more primitive dust (or dust that is activated by ice sublimation dynamics at significantly lower temperatures, not necessarily pristine). The Johnson $V$$-$$R$ color of \citep{Jewitt2015} was converted by Sloan {\sl{g}$-$$r$} following \citet{Jordi2006} and the color transformation provided by the official Sloan website\footnote{\url{http://www.sdss3.org/dr8/algorithms/sdssUBVRITransform.php}}. The converted {\sl{g}$-$$r$} color of the distant LPCs has a mean value of $\sim$0.65 $\pm$ 0.07, which is slightly higher than the median color of comets active inside the orbit of Jupiter \citep{Solontoi2012}. If this difference is real, the redder coma dust of the distant LPCs compared indicates either a higher abundance of large particles (much larger than the observation wavelengths; e.g., \citealt{Kolokolova1997,Haslebacher2024}) and/or absorbing carbonaceous materials (e.g., \citealt{KolokolovaJoc1997,Swamy2010}), the latter of which is thought to be one of the likely mechanisms behind distant comet activity \citep{Ivanova2019}.

\subsubsection{Numerical modeling of polarimetric properties} \label{sec:dis2-2}

A numerical modeling technique was employed to narrow down dust properties to reproduce the observed polarization-phase curve and polarization color of K2 dust coma by using the fast superposition T-matrix method (FaSTMM), an efficient scattering method for aggregates and clusters containing arbitrary inhomogeneous particles. Its numerical solution was computed with the method of \citet{Markkanen2017}, which has successfully explained in-situ and ground-based results of the Rosetta space mission on 67P/Churyumov-Gerasimenko (e.g., \citealt{Markkanen2018,Kwon2022}).

Given the hierarchical structure of cometary dust \citep{Guttler2019}, incident light of visible wavelength predominantly interacts electromagnetically with the dust on a $\sim$1-10-$\mu$m aggregate scale; thus, we built aggregates of equivalent size, rather than building full dust agglomerates (aggregates of aggregates). The ballistic particle cluster aggregation (BPCA) method was used to pack an aggregate in the log-normal monomer size distribution with a mean of 0.1 $\mu$m and a standard deviation of 0.048 $\mu$m. The Micro-Imaging Dust Analysis System (MIDAS) on board Rosetta measured the smallest dust units with a log-normal size distribution with similar mean and standard deviation as our model \citep{Mannel2019}. 
The resulting porosity was defined as 1 $-$ ($V_{\rm mon}$/$V_{\rm eff}$), where $V_{\rm mon}$ and $V_{\rm eff}$ are the volumes of monomers in the aggregate and a sphere with the effective radius of the aggregate, respectively. The porosity may be different for a single aggregate, but when averaged over a large number of aggregates it is around 0.7. The monomer size distribution is the same for all aggregates and the aggregate size depends on the number of monomers, where the smallest aggregates have 16 and the largest has 16,384 monomers, corresponding to the aggregates' effective radii ($\sqrt{5/3}$ * gyration radius) of 0.5$-$5 $\mu$m. For instance, an aggregate with 2048 monomers has an effective radius $r$ of 2.35 $\mu$m.

The scattering matrices $S(p)$ were averaged over a power-law size distribution of index $p$ as
\begin{equation}
    S(p) = \int_{r_{\rm min}}^{r_{\rm max}} S(r) * r^{-p} * \frac{1}{A} {\rm dr}~,
\label{eq:eq6}
\end{equation}
\noindent where $r$ is the aggregate radius and the integration limit ranges from $r_{\rm min}$ (0.5 $\mu$m) to $r_{\rm max}$ (5 $\mu$m). The normalization coefficient $A$ was defined as
\begin{equation}
    A = \int_{r_{\rm min}}^{r_{\rm max}} r^{-p} {\rm dr}~.
\label{eq:eq7}
\end{equation}
\noindent Hence, the integration contains aggregates of all sizes between 0.5 $\mu$m and 5 $\mu$m. We computed sixteen scattering elements ($S_{\rm ij}$, where $i$ and $j$ denote the number of rows and columns, respectively) of a 4$\times$4 Mueller matrix for each modeled case per each phase angle, from which we calculated the degree of linear polarization $P_{\rm r}$ of ($-$$S_{\rm 12}$ / $S_{\rm 11}$) * 100 (\%) at two wavelengths (691 nm and 834 nm). We considered six different $p$ values ranging from 2.0 to 5.0 in 0.5 increments. A real part of the refractive index ($m_{\rm r}$ for $m_{\rm r}$ + $i$$m_{\rm i}$) is fixed as 1.6, while different imaginary part values were tested for $m_{\rm i}$ = 0.01, 0.05, 0.1, 0.15, and 0.2.

It should be noted that the actual $P_{\rm r}$ value and its wavelength dependency is a non-trivial function of size and refractive index. Dust of different ice-to-silicate ratios would have largely different bulk $m_{\rm r}$ and thus produce different resulting polarization-phase curves (e.g., \citealt{Mackowski2022}). Results can also be sensitively dependent on the wavelength dependence on $m_{\rm r}$ and $m_{\rm i}$ that results from, for example, the silicate-to-organic ratios and detailed nature of absorbing materials\footnote{The effective refractive indices of several possible ingredients of cometary dust can be found at \url{https://cosmicdust.astro.umd.edu/documentation/materials}.} \citep{KolokolovaJoc1997}. In practice, we do not have a handy value for the exact stoichiometry of absorbing materials of cometary dust nor their refractive index \citep{Quirico2016}. Therefore, for the quantitative estimation in this study, we assumed (effectively) fixed $m_{\rm r}$ of 1.6 with wavelengths and considered materials' absorptivity by changing $m_{\rm i}$ only. $m_{\rm i}$ was further used in two ways to estimate the polarization color ({\it PC}), either wavelength-dependent or wavelength-independent. The former assumes the same (or effectively invariant) $m_{\rm i}$, while the latter assumes different $m_{\rm i}$ for both Red and I-filters. Given the red (positive-sloped) spectrum of K2 \citep{Meech2017,Kwon2023a} and average cometary dust \citep{Kolokolova2004,Solontoi2012}, there is a greater likelihood that $m_{\rm i}$ decreases at longer wavelengths than increases (for wavelength-dependent $m_{\rm i}$). As such, we computed the {\it PC} by setting a smaller $m_{\rm i}$ at 834 nm by $\sim$0.05 than at 691 nm (right column in Fig. \ref{Fig15}) and then compared the modeled outcomes with those of wavelength-independent $m_{\rm i}$ (left column in Fig. \ref{Fig15}). The decrement in $m_{\rm i}$ was determined arbitrarily based on the $m_{\rm i}$ option that we modeled ($m_{\rm i}$ = 0.01, 0.05, 0.1, 0.15, 0.2), rather than through exhaustive testing. A dedicated theoretical study using more realistic refractive indices and dust structure will enable better fits and more convincing evidence of dust properties.

\begin{figure*}[t!]
\centering
\includegraphics[width=0.93\textwidth]{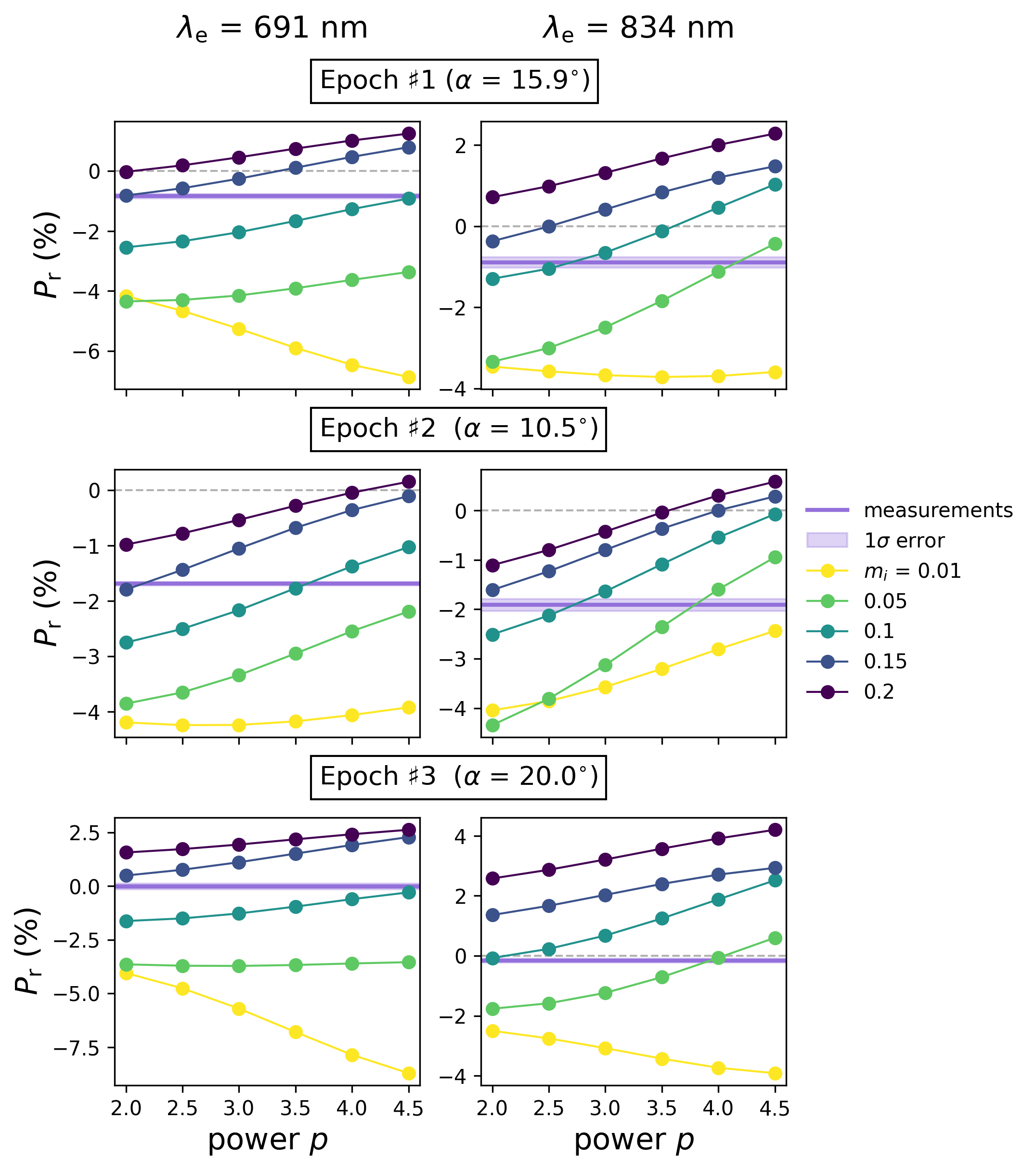}
\caption{Modeled polarization-phase curves $P_{\rm r}(\alpha)$ of cometary dust. Presented are the modeled distributions of dust particles with different power-law indices $p$ in the Red (left column) and I-filter (right column) domains. 
Each row shows observations acquired for each epoch of polarimetry. A solid horizontal line with colored areas indicates a $P_{\rm r}$ value and its 1$\sigma$ uncertainty. Each line of colored circles representing different imaginary parts of the refractive indices $m_{\rm i}$ (labeled in the legend) stamps modeled $P_{\rm r}$ at a given $p$. The case of $p$ = 5.0 is not present here due to its almost identical result to $p$ = 4.5.
\label{Fig14}}
\end{figure*}

\begin{figure}[th]
\centering
\includegraphics[width=0.73\textwidth]{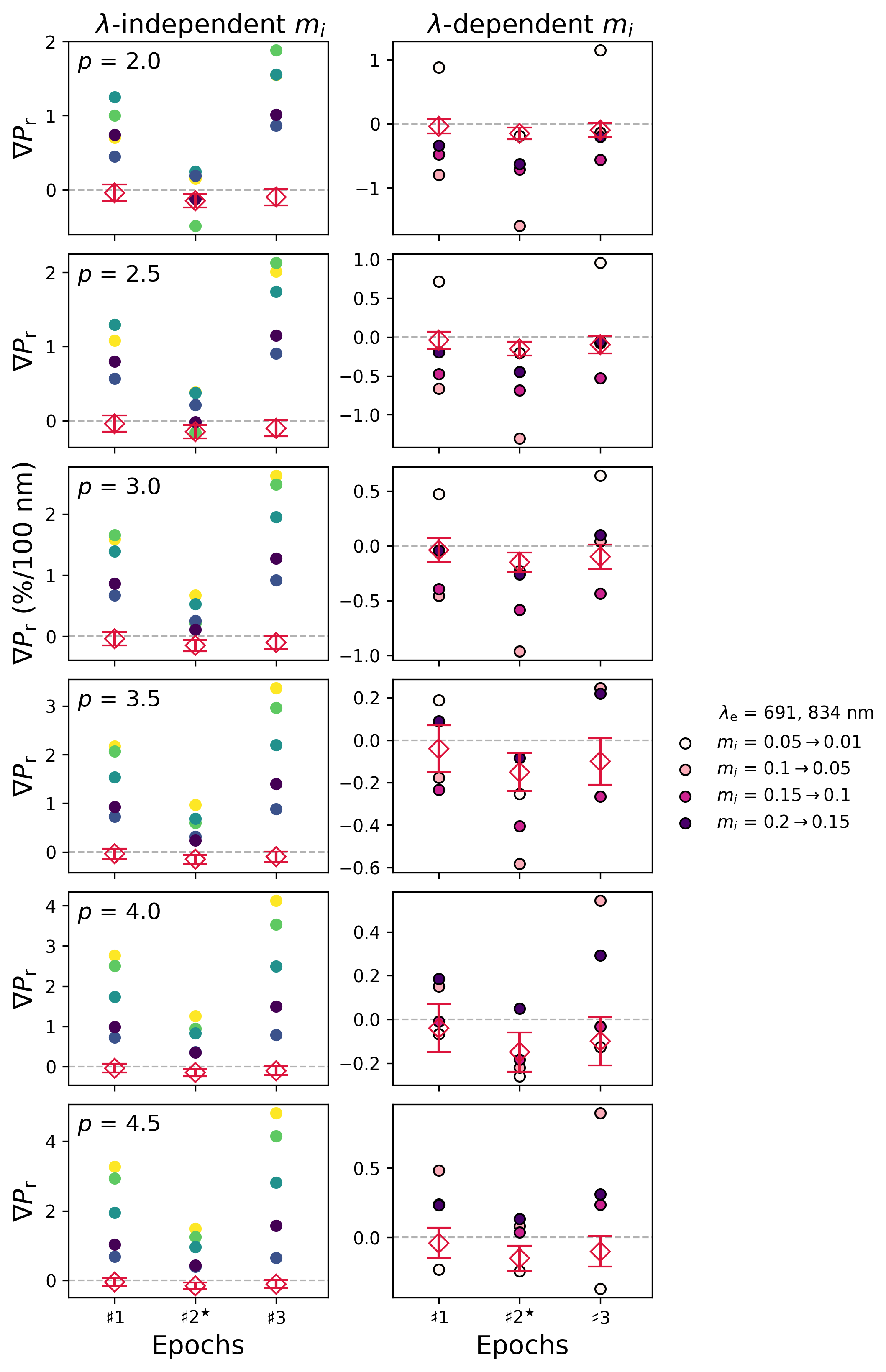}
\caption{Modeled polarization colors $P_{\rm r}(\lambda)$ of cometary dust. Columns left and right show modeled results for dust with wavelength-dependent and wavelength-independent imaginary parts of the refractive index ($m_{\rm i}$), respectively. The color scheme in the left column follows that of Figure \ref{Fig14} to distinguish cases with different $m_{\rm i}$, while the colors used in the right column are given in the legend. Each row presents PC for different $p$ values, compared with the observed results at three epochs (Red diamonds). As in Figure \ref{Fig14}, we omitted the result of $p$ = 5.0 due to its similarity to that of $p$ = 4.5.
\label{Fig15}}
\end{figure}

The modeled polarization-phase curves and {\it PC} are shown in Figures \ref{Fig14} and \ref{Fig15}, respectively, along with K2 polarimetric measurements at three epochs (Table \ref{t02}). In light of the approximate nature of the modeled dust (e.g., non-hierarchical structure) possibly reducing the accuracy of the estimation near the inversion angle (e.g., \citealt{Kolokolova2018}), direct comparison of the modeled outcomes around Epoch $\sharp$3 may be interpreted with less significance. In Figure \ref{Fig14}, the dust with very low absorptivity (i.e., $m_{\rm i}$ $\lesssim$ 0.05) or very high absorptivity ($m_{\rm i}$ $\gtrsim$ 0.2) can be immediately ruled out of viable dust properties, as none of them could explain the observed $P_{\rm r}$. This leaves either a typical or slightly more absorbing composition as a favorable option. Although this phase-curve modeling alone cannot pinpoint a unique set of dust properties, at a glance dust having $m_{\rm i}$ $\sim$ 0.1$-$0.15 in the Red domain and $m_{\rm i}$ $\sim$ 0.05$-$0.1 in the I-filter domain can reasonably well explain all-epoch measurements simultaneously. Those indices fall into the typical range of cometary dust \citep{Hadamcik2009,Levassuer-Regourd2018}, as well as consistent with the positive slope of the coma spectra of K2 over this wavelength region \citep{Meech2017,Kwon2023a}. Constraints on $p$ are not as tight as $m_{\rm i}$; nevertheless, when we assume a reasonable range of $m_{\rm i}$, the midpoint of $p$ distributes roughly around 3.0$-$3.5, which is typical of active cometary dust (e.g., \citealt{Swamy2010}). Interpreting {\it PC} of the modeled dust properties in Figure \ref{Fig15} is less straightforward than polarization-phase curves since the modeled outcomes are not a simple linear function of the parameters considered ($m_{\rm i}$ and $p$) and those trends differ between wavelength-independent $m_{\rm i}$ (left column) and wavelength-dependent $m_{\rm i}$ (right column). A quick comparison of wavelength-independent and wavelength-dependent $m_{\rm i}$ reveals that wavelength-independent $m_{\rm i}$ is less compatible with observations across all epochs. The modeled polarimetric parameters will be used in the following subsection to discuss viable dust characteristics around each epoch of polarimetric observation.
 
\subsubsection{Tentative conclusions on dust properties dominating the coma signal at each epoch} \label{sec:dis2-3}

Taking into account the available results altogether is useful to reduce ambiguity about dust particles. We combined the light-scattering measurements of K2 (i.e., polarization degree $P_{\rm r}$ in the Red and I-filter domains in Fig. \ref{Fig07}, polarization color {\it PC} $P_{\rm r}(\lambda)$ in Fig. \ref{Fig09}, {\sl{g}$-$$r$} coma color $m_g$$-$$m_r$ in Fig. \ref{Fig13}, and coma morphology), in conjunction with the modeling results in Figures \ref{Fig14} and \ref{Fig15}, to propose the most viable dust characteristics dominating each epoch's coma signal.

\paragraph{Epoch $\sharp$1 ($r_{\rm H}$ = 3.23 au \& $\alpha$ = 15.9\arcdeg)}~
All the polarimetric measurements at this epoch are typical of observations of active comets (Figs. \ref{Fig07} and \ref{Fig09}).  In Figure \ref{Fig14}, $P_{\rm r}$ in the Red domain shows its best compatibility with the dust of $m_{\rm i}$ $\sim$ 0.1$-$0.15 and $p$ $\sim$ 3.0$-$3.5, but a wider option of $p$ seems also possible within the typical composition range. 
$P_{\rm r}$ in the I-filter domain has a preference in the dust of $m_{\rm i}$ $\sim$ 0.1 but favors slightly smaller $m_{\rm i}$ values than in the Red domain. Dust having $m_{\rm i}$ of $\lesssim$0.05 is incompatible with the measured $P_{\rm r}$ at both wavelengths, let alone with the $m_g$$-$$m_r$ colors falling into the typical range of cometary dust \citep{Levassuer-Regourd2018}. {\it PC} results show that none of the modeled cases are viable if $m_{\rm i}$ is effectively invariant with wavelength (left column in Fig. \ref{Fig15}), which seems reasonable given the above conjecture made from the modeled $P_{\rm r}$ cases. By comparison, a wavelength-dependent $m_{\rm i}$ (right column in Fig. \ref{Fig15}) could fit the measured {\it PC} with a $p$ $\sim$ 3.0$-$4.0, but it is difficult to pinpoint a unique solution due to the limited dust approximation pointed out above. We conclude that the K2's coma at Epoch $\sharp$1 would be describable with the dust having $m_{\rm i}$ of $\sim$0.1 and $p$ of $\sim$3.0, all of which are typical cometary dust parameters.

\paragraph{Epoch $\sharp$2 ($r_{\rm H}$ = 2.78 au \& $\alpha$ = 10.5\arcdeg)}~$P_{\rm r}$ of both filters require $m_{\rm i}$ $\gtrsim$ 0.05 in Figure \ref{Fig14}. Compared to the Epoch $\sharp$1 data, dust having slightly smaller $m_{\rm i}$ and/or smaller $p$ on average provides a better fit. If we assume nearly wavelength invariant $m_{\rm i}$ over the two wavelength regimes, only dust with $p$ = 2.0$-$2.5 can match with the derived {\it PC} (left column in Fig. \ref{Fig15}). In the case of wavelength-dependent $m_{\rm i}$ (right column in Fig. \ref{Fig15}), intermediate $p$ cases ($p$ $\sim$ 3.0$-$4.0) are not impossible to match with the observations, although a practical interpretation is difficult due to the nonlinear dependence of {\it PC} on $m_{\rm i}$ and $p$. This difficulty may be due to the complexity of modeling realistic dust particles and also the possibility of inappropriate $m_{\rm i}$ decrements for the dust approximation we tested (likely too-steep $m_{\rm i}$ change over the filter domains examined; \citealt{KolokolovaJoc1997}). Despite this, the observed phase angle and wavelength dependencies of $P_{\rm r}$ do not necessarily indicate a substantial difference in the dust properties from Epoch $\sharp$1; thus, we conclude that coma dust roughly similar to Epoch $\sharp$1 ($m_{\rm i}$ of $\sim$0.1 and $p$ of $\sim$3.0, preferably slightly smaller $m_{\rm i}$ and/or smaller $p$) can fit Epoch $\sharp$2 data reasonably.

\paragraph{Epoch $\sharp$3 ($r_{\rm H}$ = 2.53 au \& $\alpha$ = 20.0\arcdeg)}~At this epoch, a wavelength-independent $m_{\rm i}$ of $\sim$0.1$-$0.15 in most of the modeled $p$ cases appear to provide a reasonable fit to the measured {\it PC} (left column in Fig. \ref{Fig15}). Dust of $p$ = 4.5 for wavelength-dependent $m_{\rm i}$ might be plausible, yet observed $P_{\rm r}$ in both filter domains in Figure \ref{Fig14} and typical $m_g$$-$$m_r$ coma colors in Figure \ref{Fig13} prefer rather broadly similar $m_{\rm i}$ and $p$ to those at Epoch $\sharp$1 ($m_{\rm i}$ of $\sim$0.1$-$0.15 and $p$ of $\sim$3.0$-$4.0), rejecting such extreme cases of the parameters. We conclude that the inner coma dust at this epoch, as observed around the inner boundary of the water ice line, seems broadly consistent with Epoch $\sharp$1 and Epoch $\sharp$2.

\vspace*{+.1in}
Our observations and simplified dust model provide limited insight into K2's dust properties, but all of our estimates strongly suggest that K2's secular change in dust particle characteristics dominating its real-time inner (an order of $\sim$10$^3$ km) coma around the sublimation boundary of water does not deviate significantly from the trend expected of active comets. By comparison, all the discontinuities observed in large-scale (an order of $\sim$10$^4$ km) $r$-band brightness, {\it Af$\rho$}, coma morphology, and coma color indicate that the comet certainly underwent discontinuous activation around 2.9 au from the Sun. All of these measurements on K2, including the highlighted variations in different coma radial scales, fall within the broad range of dust characteristics for active comets in the inner solar system (\citealt{Kolokolova2004,Levassuer-Regourd2018} for a review).\\

\subsection{Secular evolution of the preperihelion activity of K2} \label{sec:dis3}

By combining multi-band monitoring photometric data covering different ice-sublimation regimes with imaging polarimetric data collected around the passage of water-ice sublimation-driven boundary and dust modeling implications, we attempted to discuss the plausible nucleus environment of K2 within a broader evolutionary context.

The heliocentric evolution of the coma brightness and dust production rate (represented by {\it Af$\rho$}) of K2 (Fig. \ref{Fig01}) reveals significant activity exhibiting $\sim$10$-$1,000 times the value of typical active comets of similar geometry \citep{Licandro2019,Garcia2020}. Such substantial activity in the supervolatile ice-driven regime (primarily CO at $r_{\rm H}$ $\gtrsim$ 6$-$7 au; e.g., \citealt{Womack2017}) indicates the comet's CO abundance in its nucleus, which is also supported by a flux excess observed on UT 2021 March 31 at $r_{\rm H}$ = 6.735 au on the NEOWISE W2 band (effective wavelength of 4.6 $\mu$m) that includes CO$+$CO$_2$ emission bands (Appendix \ref{sec:app2}). An appreciable CO detection would also be compatible with a radio wavelength measurement of CO molecules in the sunward coma at $r_{\rm H}$ = 6.72 au \citep{Yang2021}. Such an exceptional distant activity indicates K2's nucleus is accommodating abundant supervolatile. It may be due to K2's early ejection from the Oort cloud \citep{Lisse2022,Anderson2022} and/or its current dynamics (long orbital period of $\sim$3 Myr and high eccentricity of $e$ = 0.9998) that keep a primitive surface environment that results in its current activity being a direct response to simultaneous Sun heating \citep{Krolikowska2018}. 
However, K2's power-law heliocentric dependence of {\it Af$\rho$} with an index of $\sim$0.9 (Fig. \ref{Fig01}, consistent with the estimation by \citealt{Fulle2022}) in this heliocentric regime ($r_{\rm H}$ $>$ 6 au) implies that the cometary surface may not be perfectly pristine. Under the scheme of a comet nucleus made up of centimeter-sized pebbles and its activity driven by ice embedded in the pebbles \citep{Blum2018,Fulle2019}, \citet{Fulle2022} suggest in their Eq. 9 that a linear relationship between dust mass loss rate and {\it Af$\rho$} allows for evaluation of the significance of fallback dust on the comet nucleus by observing a $r_{\rm H}$ dependence of {\it Af$\rho$}. Assuming both the nucleus erosion rate and dust ejection terminal velocity depend on $r_{\rm H}^{-1}$ in their coma models (e.g., \citealt{Zakharov2018}), {\it Af$\rho$} varying close to $r_{\rm H}^{-1}$ indicates that the observed dust mass loss rate varies more steeply with $r_{\rm H}$ than predicted by the theoretical nucleus erosion rate, which implies significant dust particle fallback onto the cometary surface \citep{Fulle2022}. This coma dynamics inferred seem to differ from those of distant comets with perihelion distances less than 3 au, at similar $r_{\rm H}$ in their inbound legs, whose coma dust particles are typically micron-sized and made up of tholin-like absorptive materials (e.g., \citealt{Ivanova2019}).

Two polarimetric studies on the K2 coma dust were conducted in this distant solar system regime, at $r_{\rm H}$ of $\sim$7.14--6.80 au ($\alpha$ of 7.39--8.39$^\circ$) by \citet{Kochergin2023} and $\sim$6.82 au ($\alpha$ of 8.35$^\circ$) by \citet{Zhang2022}. Our results on K2 made in the inner solar system are compatible with these studies, in the sense that 1) the $P_{\rm r}$ values of the coma are similar to those from other active comets at similar phases, and 2) the $P_{\rm r}$ values for different aperture sizes vary slightly, but are generally constant within two standard deviations (with $\rho$ of 10,000 and 20,000 km; \citealt{Kochergin2023}). Both groups, however, observed a more dynamic dust coma environment than our study, suggesting the presence of micron-sized transparent dust particles, possibly water ice mixed with carbonaceous materials, sublimated in the inner coma region, implying active comet status in the presence of dust particles released from the nucleus. This active sublimation would be responsible for the observed $P_{\rm r}$ radial gradient outward $\rho$ $\sim$ 20,000 km (Fig. 2 in \citealt{Zhang2022}).

Interestingly, over $r_{\rm H}$ of 4$-$6 au between the first and second inflection points marked by triangles in Fig. \ref{Fig01}, where CO$_2$ ice is likely to lead cometary activity \citep{Fulle2022}, the $r$-band absolute magnitude remained almost constant or rather slightly decreased. The {\it Af$\rho$} also gradually declined until $\sim$2.9 au, suggesting a slowdown in brightness in this heliocentric regime. Direct observation of CO$_2$ emissions is impossible to make using ground-based telescopes; nonetheless, recent preperihelion studies on K2, such as endorsing CO$_2$ as the main parent source of atomic oxygen emission lines [OI] distributing over $\sim$558$-$636 nm \citep{Feldman2004} observed at $r_{\rm H}$ of $\sim$2.8 au \citep{Cambianica2023}, have supported at least the existence of volatile ice driving the dust ejection. It is unclear for now whether the apparent decrease in brightness in this activity regime is related to the comet's underabundance in CO$_{\rm 2}$ or low CO$_{\rm 2}$/CO ice ratio. We expect that infrared spectroscopy results on K2 from the James Webb Space Telescope will allow an in-depth discussion of the ice composition \citep{Woodward2023,Wooden2023}.

As a final phase, K2 entered the realm whose temperature is warm enough to ignite water ice sublimation ($r_{\rm H}$ $\sim$ 4 au; \citealt{Fulle2020,Ciarniello2022}). Our Epoch $\sharp$1 polarimetric data at $r_{\rm H}$ of $\sim$3.2 au shows coma dust that is typical of preperihelion short-period comets \citep{Hadamcik2009,Snodgrass2016,Levassuer-Regourd2018}), akin to the dust consisting of the steady-state K2 coma observed outside Saturn \citep{Jewitt2017a,Meech2017,Hui2017}. The old, stale coma dust would be replaced when the comet crossed at $\sim$2.85 au (Epoch $\sharp$2), where the onset of water-ice sublimation was evident, yielding powerful cometary activity. A solar insolation-driven dust plume was likely launched, ejecting dust in a wide size distribution. Our polarimetric data would capture large dust chunks lingering in the nucleus dominate the inner coma component, whereas concurrent ZTF multiband photometry data explore the outer coma region that is subject to the signal of smaller, higher-mobility dust particles and C$_2$-like photo-dissociated gas molecules. Meanwhile, water ice chunks likely embedded in previously ejected large dust particles in the CO-ice sublimation distance were activated and showed sublimation evidence \citep{Kwon2023a}. Such water-ice sublimation-driven activity might not be compatible with the apparent absence of H$\alpha$ emissions measured in our narrowband photometry (Sect. \ref{sec:res4}). However, given the lack of monitoring data of H$\alpha$ as well as other gas emissions having affinity with water ice covering a smaller heliocentric regime and the presence of prominent peaks of forbidden oxygen lines [OI] on K2 at similar $r_{\rm H}$ \citep{Kwon2023a}, the H$\alpha$ measurements in this study may hardly provide conclusive evidence on the intrinsic abundance of water ice on K2.

Our ZTF data covered before and after the activation at $\sim$2.9 au, one on UT 2022 June 09.3 at $r_{\rm H}$ $\sim$ 3.0 au and the other on UT 2022 June 17.3 at $r_{\rm H}$ $\sim$ 2.8 au. Two epochs' apparent $r$-band magnitudes $m_{\rm r,app}$ measured with $\rho$ = 20,000 km were converted to the absolute magnitude $H_{\rm r}$ using Eq. \ref{eq:eq0}. The derived $H_{\rm r}$ is then used to estimate the effective cross-section of the scattering dust cloud $C_{\rm d}$ in the same manner as \citet{Russell1916,Jewitt2018} using
\begin{equation}
    p_{\rm r}C_{\rm d} = 2.25 \times 10^{22} \pi 10^{-0.4(H_{\rm r}-m_{\rm \sun,r})}~,
\label{eq:eq9}
\end{equation}
\noindent $p_{\rm r}$ is the dust's geometric albedo. Assuming $p_{\rm r}$ = 0.04, $C_{\rm d}$ increases by $\delta C_{\rm d}$ $\sim$ 5.5 $\times$ 10$^7$ m$^2$ between the two observations. Based on our results and previous polarimetric studies (e.g., \citealt{Kolokolova2007}), if we assume the coma dust aggregate having a power-law size distribution dominated by large aggregates in the optical (the effective size of dust $a_{\rm d}$ $\sim$ 5 $\mu$m, which is the largest dust among the considered by our numerical calculation, assuming that the large end of the particle size distribution dominates the mass of the dust coma; e.g., \citealt{Lisse2002,Green2004,Marschall2020a}) with its sphericity and average bulk density $\rho_{\rm d}$ of 785 kg m$^{-3}$ \citep{Fulle2017}, the average dust cloud mass production rate (despite the high likelihood of abrupt, rather than constant, ejection) within $\rho$ = 20,000 km equivalent to the increase of the activity $\delta M_{\rm d}$/$\delta$t would be 4$\rho_{\rm d}$$a_{\rm d}$$\delta C_{\rm d}$/(3$\tau_{\rm d}$) $\sim$ 0.42 kg s$^{-1}$, where $\tau_{\rm d}$ is the travel time of the ejected dust particles during the two terms ($\sim$8 days = 691,200 secs). 
Assuming the spherical nucleus of K2 having a effective radius $r_{\rm nuc}$ of 3 km \citep{Jewitt2017b,Fulle2022} and bulk nucleus density $\rho_{\rm nuc}$ of $\sim$500 kg m$^{-3}$ (e.g., \citealt{A'Hearn2011,Patzold2019}), ejected dust mass $\delta M_{\rm d}$ over the two terms accounts for well below 1 \% of the mass of the nucleus. 
The dust loss calculated using the visible-light data here would be significantly underestimated (by several orders of magnitude) compared to that calculated using more plausible ejection of big particles ($a_{\rm d}$ $\sim$ 100 $\mu$m$-$1 cm).

The new coma component responsible for accelerating brightening began governing the coma signal once the comet reached $<$2.7 au, which was covered by our Epoch $\sharp$3 polarimetric data. All observational parameters during this term fall within the typical range of active solar system comets that have been observed under similar conditions.
\\

\section{Summary} \label{sec:sum}

In this paper, we report new VLT/FORS2 imaging polarimetric observations of active Oort-cloud comet C/2017 K2 (PANSTARRS) as it approached the Sun and particularly passed through its water-ice sublimation boundary around 2.7 au from the Sun (visited at three epochs, each covering before, during, and after the passage of the region), aided by archival multi-band ZTF and NEOWISE data. The main results and discussion points are summarized below:

\begin{enumerate}
\item[$\bullet$] Over the passage of the water ice line, K2 displays homogeneous radial distributions in the coma in its $r$-band intensity and Red- and I-filter domain polarization degrees $P_{\rm r}$ within the 2$\sigma$ error bars, indicating that the comet apparently lacks significant anisotropic coma features. 

\item[$\bullet$] Aperture-averaged over $\rho$ $\sim$ 1,300$-$10,000 km, a phase-angle dependence of polarization degree $P_{\rm r}(\alpha)$ of the dust coma of K2 was compared to the trends of other comets. We obtained $P_{\rm r}$ = $-$0.84 $\pm$ 0.07 \% in the Red domain and $-$0.89 $\pm$ 0.13 \% in the I-filter domain at $\alpha$ = 15.9\arcdeg (Epoch $\sharp$1); $P_{\rm r}$ = $-$1.69 $\pm$ 0.03 \% in the Red domain and $-$1.91 $\pm$ 0.12 \% in the I-filter domain at $\alpha$ = 10.5\arcdeg (Epoch $\sharp$2); and $P_{\rm r}$ = $-$0.01 $\pm$ 0.14 \% in the Red domain and $-$0.16 $\pm$ 0.06 \% in the I-filter domain at $\alpha$ = 20.0\arcdeg (Epoch $\sharp$3).
The polarimetric measurements of K2 over all epochs are well consistent with those of typical active comets observed in the NPB.

\item[$\bullet$] A spectral dependence of polarization $P_{\rm r}(\lambda)$, or polarization color {\it PC}, of the K2 dust coma covering the inner coma component was derived as $-$0.04 $\pm$ 0.11 \% per 100 nm at Epoch $\sharp$1, $-$0.91 $\pm$ 0.12 \% per 100 nm at Epoch $\sharp$2, and $-$0.16 $\pm$ 0.06 \% per 100 nm at Epoch $\sharp$3, all of which are well within the deviation of the measured values for active comets at similar phases. 

\item[$\bullet$] Preperihelion activity of K2 with ZTF zr-band data over $r_{\rm H}$ $\sim$ 14$-$2.3 au showed two inflection points at $\sim$6 and $\sim$2.9 au where visible activity changes its evolution tendency. At $r_{\rm H}$ $\gtrsim$ 6$-$7 au, K2's $Af\rho$ parameter increased with a power-law index of roughly $-$0.9 as $r_{\rm H}$ decreases, implying the significance of the fall-back dust ejection and thus supporting the past cometary activity in the distant solar system \citep{Fulle2022}. The NEOWISE W2 band excess at $r_{\rm H}$ = 6.735 au supports the activity in this regime. Over $r_{\rm H}$ = 4$-$6 au, the rate of increase in $r$-band magnitude stagnated or slightly decreased. Once the comet enters $r_{\rm H}$ $<$ 3 au, the inner coma separates its orientation along the Sun$-$comet radial direction from the outermost coma. 

\item[$\bullet$] The 20,000-km-wide coma dust at preperihelion covered by ZTF exhibited $g-r$ color nearly consistent with the 1$\sigma$ distribution of average solar system comets between $r_{\rm H}$ $\sim$3.3 and 2.4 au (0.57 $\pm$ 0.05; \citealt{Solontoi2012}). However, around 2.9 au where the comet's brightening trend shows discontinuities (evidence for this can be seen in the $r$-band magnitude and {\it Af$\rho$}), both inner (5,000 km from the center) and outer coma colors temporarily experienced abrupt bluing and subsequent reddening, followed by returning to their pre-event conditions at $\sim$2.5 au.

\item[$\bullet$] H$\alpha$-band emission at 656.3 nm has not enough signals for photometry in all of the VLT imaging observations in line with the previous studies sharing the negligible signal of the same band in the VLT/MUSE spectro-photometry at $r_{\rm H}$ $\sim$ 2.5 au \citep{Kwon2023a}.

\item[$\bullet$] We used FaSTMM numerical modeling of dust in the inner coma to suggest that dust with power-law size distribution with an index $p$ $\sim$ 3.0$-$3.5 and an absorbing composition (imaginary refractive index $m_{\rm i}$ $\sim$ 0.1$-$0.15, possibly wavelength-dependent decreasing at longer optical wavelengths), typical of the dust of active comets \citep{Kolokolova2004,Levassuer-Regourd2018}, could explain the polarimetric and color properties observed on K2 during the crossover of the water ice line. 

\end{enumerate}

Monitoring the secular evolution of the perihelion cometary activity over a wide range of heliocentric distances, particularly covering the boundaries of activity regimes where the dominant activity leads changes from one to another type of ice, is of utmost importance to offer a comprehensive understanding of the coma environment and thus the global scale of the (near-)surface layer of the comet nucleus. Future studies aided by the international collaboration of telescopic networks for long-term observations will be vital for putting comet observations into context. 

\section{Acknowledgments}
Based on observations made with ESO Telescopes at the Paranal Observatory under program 109.23D6.001 (PI: Y. G. Kwon). J.M. acknowledges the German Research Foundation DFG grant No. 517146316. This work used the supercomputer Phoenix and was supported by the Gau$\beta$-IT-Zentrum of the University of Braunschweig (GITZ). J.A. gratefully acknowledges funding from the Volkswagen Foundation. J.A. acknowledges funding from the European Union’s Horizon 2020 research and innovation program under grant agreement No. 75390 CAstRA. Research by Z. G. is funded by the UK Science and Technology Facilities Council (STFC).

Based on observations obtained with the Samuel Oschin 48-inch Telescope at the Palomar Observatory as part of the Zwicky Transient Facility project. ZTF is supported by the National Science Foundation under Grant No. AST-1440341 and a collaboration including Caltech, IPAC, the Weizmann Institute for Science, the Oskar Klein Center at Stockholm University, the University of Maryland, the University of Washington, Deutsches Elektronen-Synchrotron and Humboldt University, Los Alamos National Laboratories, the TANGO Consortium of Taiwan, the University of Wisconsin at Milwaukee, and Lawrence Berkeley National Laboratories. Operations are conducted by COO, IPAC, and UW.
This publication also makes use of data products from the Widefield Infrared Survey Explorer, which is a joint project of the University of California, Los Angeles, and the Jet Propulsion Laboratory/California Institute of Technology, funded by the National Aeronautics and Space Administration. This publication also makes use of data products from NEOWISE, which is a project of the Jet Propulsion Laboratory/California Institute of Technology, funded by the Planetary Science Division of the National Aeronautics and Space Administration. This research has made use of the NASA/IPAC Infrared Science Archive, which is funded by the National Aeronautics and Space Administration and operated by the California Institute of Technology.

%

\vspace{5mm}
\facilities{VLT:Antu (FORS2), ZTF, NEOWISE}


\software{NumPy \citep{Harris2020}, SciPy \citep{Virtanen2020}, Matplotlib \citep{Hunter2007}, astropy \citep{astropycollab2022},  lacosmic \citep{vanDokkum2001}, photutils \citep{Bradley2023}}




\appendix
\counterwithin{figure}{section}
\section{Quality checks} \label{sec:app1}

\begin{figure}[!b]
\centering
\includegraphics[width=\textwidth]{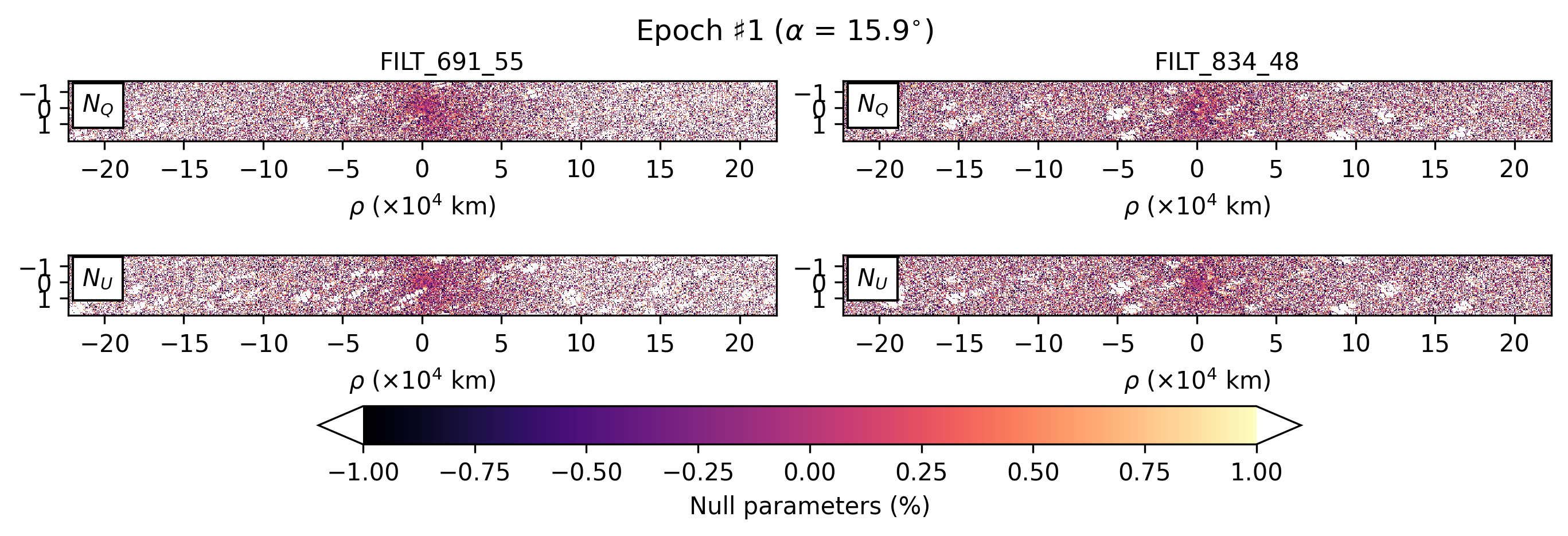}
\caption{Maps of null parameters $N_{Q}$ and $N_{U}$ of the first-epoch imaging polarimetric data taken in FILT\_691\_55 (left) and FILT\_834\_48 (right) filters. Comet is located in the center of the maps. Any values outside the range of $\pm$1 \% appear white.}
\label{Figap1}
\end{figure}

\begin{figure}[ht]
\centering
\includegraphics[width=\textwidth]{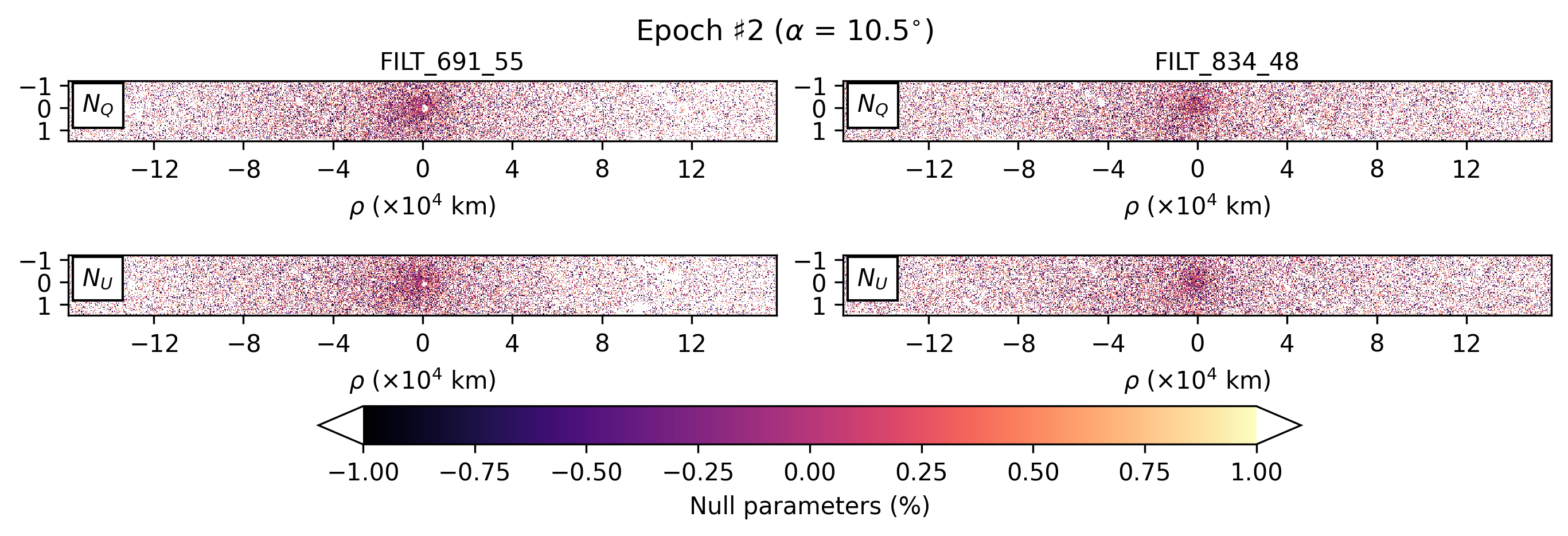}
\caption{Same as Figure \ref{Figap1} but for the second-epoch data.}
\label{Figap2}
\end{figure}

\begin{figure}[ht]
\centering
\includegraphics[width=\textwidth]{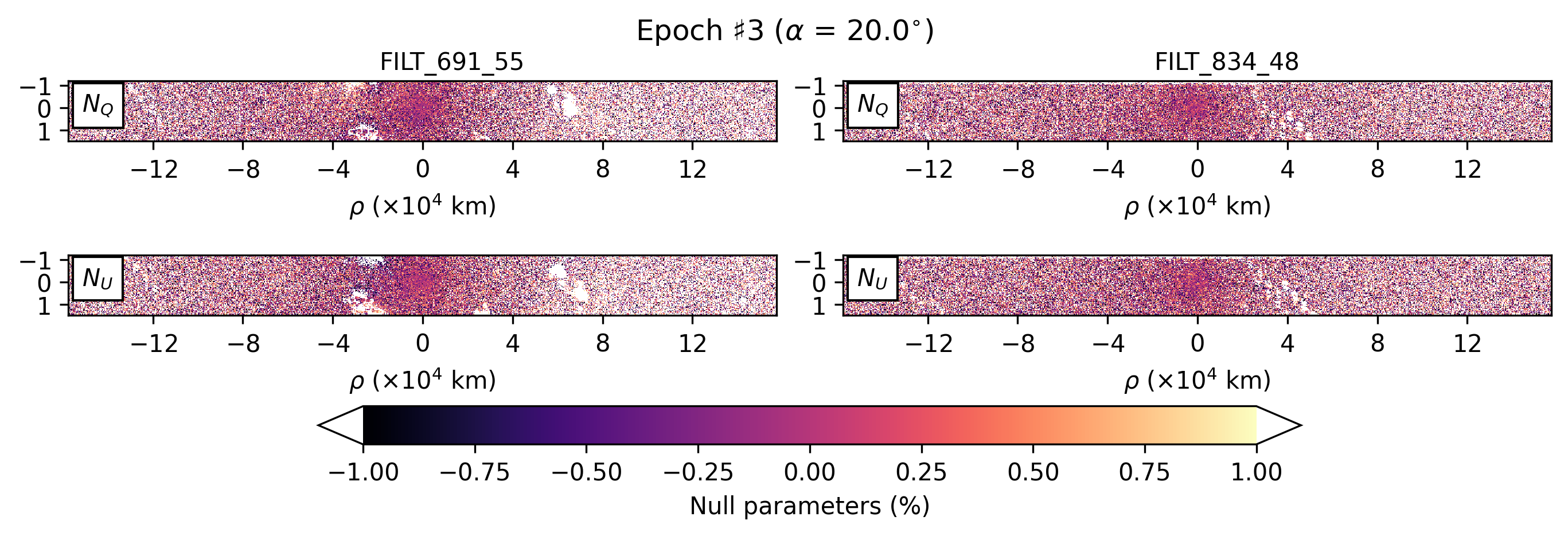}
\caption{Same as Figure \ref{Figap1} but for the third-epoch data.}
\label{Figap3}
\end{figure}

To verify the validity of derived polarization degrees $P_{\rm r}$ and their spatial distribution, we created maps of their null parameters $N_{Q}$ and $N_{U}$ in the same way as \citet{Bagnulo2023}. The quantities are defined on our two sets of exposure pairs of Stokes parameters as
\begin{equation}
\begin{split}
N_{Q} & = \frac{1}{4} \Bigg\{\Bigg(\frac{f_{o} - f_{e}}{f_{o} + f_{e}}\Bigg)_{0\arcdeg} - \Bigg(\frac{f_{o} - f_{e}}{f_{o} + f_{e}}\Bigg)_{45\arcdeg} - \Bigg(\frac{f_{o} - f_{e}}{f_{o} + f_{e}}\Bigg)_{90\arcdeg} + \Bigg(\frac{f_{o} - f_{e}}{f_{o} + f_{e}}\Bigg)_{135\arcdeg}\Bigg\}\\
N_{U} & = \frac{1}{4} \Bigg\{\Bigg(\frac{f_{o} - f_{e}}{f_{o} + f_{e}}\Bigg)_{22\fdg5} - \Bigg(\frac{f_{o} - f_{e}}{f_{o} + f_{e}}\Bigg)_{67\fdg5} - \Bigg(\frac{f_{o} - f_{e}}{f_{o} + f_{e}}\Bigg)_{112\fdg5} + \Bigg(\frac{f_{o} - f_{e}}{f_{o} + f_{e}}\Bigg)_{157\fdg5}\Bigg\}
\end{split}
\label{eq:ap1}
\end{equation}
These null parameters should ideally have a Gaussian distribution centered around zero with the same Full Width at Half Maximum (FWHM) as corresponding Stokes parameters.
Significant deviations from zero (greater than $\pm$1 \% when normalized by the uncertainties of the data) could indicate an issue with instruments (e.g., optics) and/or external factors (e.g., our data reduction process and cosmic rays). Null parameters, together with $P_{U}$ (Eq. \ref{eq:eq1}), and their distributions across the coma were therefore used as the final quality check tool of our measurements. Figures \ref{Figap1}, \ref{Figap2}, and \ref{Figap3} show null-parameter maps at the first, second, and third epochs, respectively.
\\


\section{Distribution of polarimetric uncertainties} \label{sec:app1-1}

\begin{figure}[h]
\centering
\includegraphics[width=0.77\textwidth]{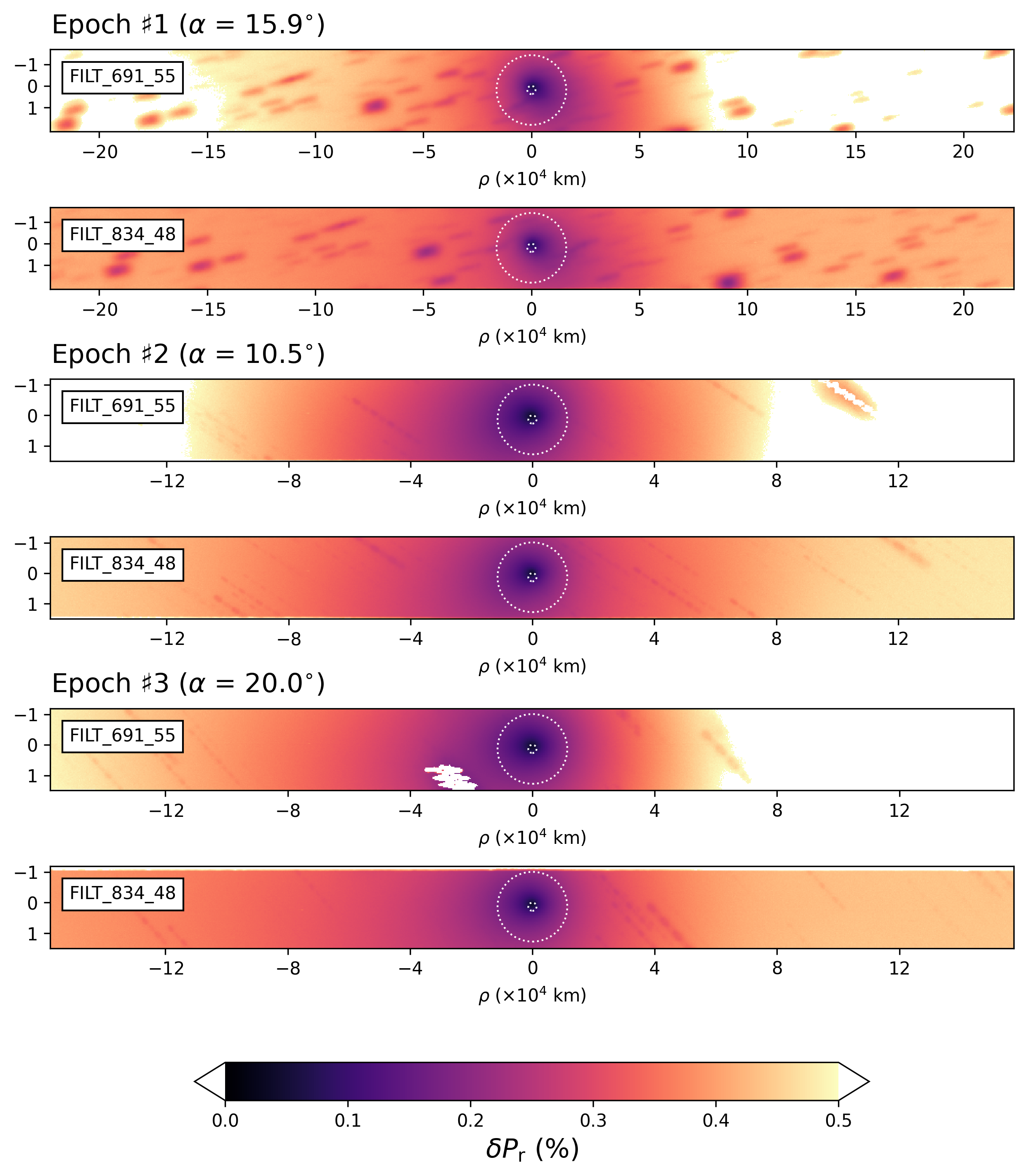}
\caption{Maps of polarimetric uncertainties of K2. Each row corresponds to the ones in Figure \ref{Fig03}, where the photocenter of the coma is located at the center (0, 0). Smaller and larger dotted circles in each panel indicate the inner and outer coma boundaries of interest.}
\label{Figap1-1}
\end{figure}

Figure \ref{Figap1-1} illustrates the distribution of uncertainties $\delta P_{\rm r}$ in the corresponding $P_{\rm r}$ maps (Fig. \ref{Fig03}).
The dotted lines indicate the inner and outer boundaries of the radial bins in which $P_{\rm r}$ values at each annulus were measured (Figs. \ref{Fig04}$-$\ref{Fig06}). The photocenter of the comet is parked at the center of each panel (0, 0) and  $\delta P_{\rm r}$ $>$ 0.5 \% areas were flagged as NaNs.

\section{CO activity regime covered by NEOWISE data} \label{sec:app2}

\begin{figure}[b!]
\centering
\includegraphics[width=0.6\textwidth]{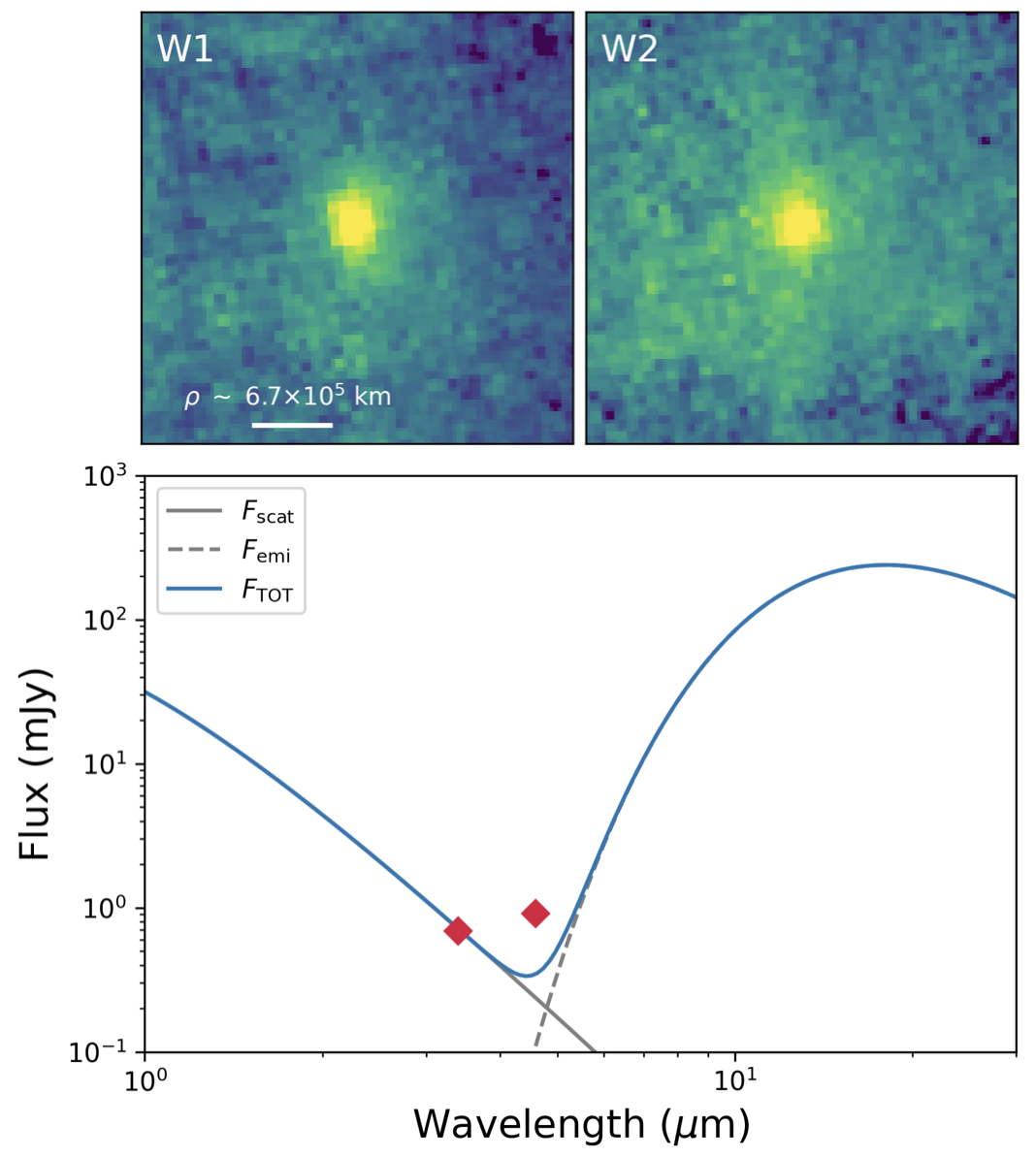}
\caption{Median-combined W1 and W2 band coma images of K2 (upper panels) and the spectral distribution of their measured fluxes (lower panel). North is to the up and east is to the left.}
\label{Figap4}
\end{figure}

The {\it Near-Earth Object Wide-field Infrared Survey Explorer} (NEOWISE) reactivation mission is a two-band infrared survey of the entire sky, aiming to detect and characterize the physical properties of small bodies in the solar system \citep{Mainzer2011,Mainzer2014}. K2 has been in the frame since the full cryogenic mission phase at $r_{\rm H}$ $\sim$ 29 au and already showed appreciable coma signals around $r_{\rm H}$ $\sim$ 16 au \citep{Meech2017}. In this study, we analyzed the K2 data taken in the CO-driven activity regime ($r_{\rm H}$ $\gtrsim$ 6 au; Fig. \ref{Fig01}) for a quick confirmation of the presence of CO/CO$_{\rm 2}$ sublimation by measuring the excess of the W2 band flux. We made use of the W1- and W2-band data median-combined of 72 fits files with the \texttt{ICORE} software \citep{Masci2009} with a mid-frame $r_{\rm H}$ = 6.735 au, $\Delta$ = 6.674 au, and $\alpha$ = 8.5\arcdeg\ on UT 2021 March 31, measured their fluxes with an aperture radius of 11\arcsec\ (corresponding to $\sim$8 pixels), and estimated their uncertainties using ancillary files of coadd standard deviation and uncertainties. As a result, the W1 band flux is 0.69 $\pm$ 0.05 mJy and the W2 band flux is 0.91 $\pm$ 0.02 mJy, whose spectral distribution is shown in Figure \ref{Figap4}.

The coma of K2 appears to be spherical both in the W1 and W2 bands. Near-infrared spectra of small bodies ($\sim$1$-$5 au) are a combination of the scattered light ($\lambda$ $\lesssim$ 2.5 $\mu$m) and thermal emission ($\lambda$ $>$ 3 $\mu$m), whose boundaries of dominance depend on $r_{\rm H}$ and physical properties of the bodies. As spectra up to wavelengths of $\sim$5 $\mu$m obtained beyond the orbit of Jupiter are generally considered to well approximate their scattered-light portion (e.g., \citealt{Emery2006}) and the W2 band contains the CO$+$CO$_{2}$ emission bands \citep{Bauer2015}, we used the W1 band flux as a reference point of the reflected dust continuum and arbitrarily adjusted the Planck spectrum of the Sun (effective temperature of 5,770 K) to the point ($F_{\rm scat}$ in Fig. \ref{Figap4}). The thermal signal was extrapolated using a modified blackbody function for coma dust with a visible albedo of $\sim$0.04 and an emissivity of 0.9 ($F_{\rm emi}$ in the same figure). In the absence of W3 and W4 data, we plotted the thermal portion of the spectrum, assuming that the dust coma of K2 has a superheat ($S$ = T$_{\rm obs}$/T$_{\rm BB}$) of 1.5 (typical for active comets albeit most of the mid-infrared observations have been made in the much inner solar system than the K2 observation here; e.g., \citealt{Hanner2004}) relative to the blackbody temperature (T$_{\rm BB}$ = 278 K / $\sqrt{r_{\rm H}}$; \citealt{Gehrz1992}) at the time of observation. Given the total emission ($F_{\rm TOT}$ = $F_{\rm scat}$ $+$ $F_{\rm emi}$), the W2 band flux always shows the excess independently of the assumed $S$ value between 1 and 3.5, compatible with the observed supervolatile-driven activity well beyond the water ice line in Figure \ref{Fig01}. Measured flux values and the W2 band excess are consistent with the previous report \citep{Milewski2024}.

\end{document}